\documentclass[a4paper,11pt]{article}
\usepackage[T1]{fontenc}
\usepackage{amsfonts}
\usepackage[english]{babel}
\usepackage{a4wide,times}

\usepackage[pdftex]{graphicx}
\usepackage{glossaries}
\makeglossaries
\usepackage[usenames,dvipsnames]{xcolor}
\usepackage[colorlinks=true, linkcolor=blue, citecolor=orange]{hyperref}
\usepackage{fancyhdr}
\usepackage{sectsty}
\usepackage{amssymb}
\usepackage{graphics}
\usepackage{epsfig}
\usepackage{amsbsy}
\usepackage{verbatim}
\usepackage{color}
\usepackage{mathrsfs}
\usepackage{makeidx}
\usepackage{amsmath}
\usepackage{amsthm}
\usepackage{enumitem}

\theoremstyle{plain}
\newtheorem{theorem}{Theorem}[section]
\newtheorem{corollary}[theorem]{Corollary}

\newtheorem{proposition}[theorem]{Proposition}

\theoremstyle{definition}
\newtheorem{definition}{Definition}[section]
\usepackage{dsfont}	

\theoremstyle{remark}
\newtheorem{remark}{\bf Remark}[section]
\theoremstyle{remark}
\newtheorem{example}{\bf Example}
\newtheorem{com*}{\bf Comment}

\usepackage{fancyhdr}

\usepackage{geometry}
\title{ Fair profit sharing ratios of Islamic investment contracts}

\author{  
{\sc  Abass Sagna  } \thanks{ENSIIE \& labouratoire de Math\'ematiques et Mod\'elisation d'Evry (LaMME), Universit\'e  \'Evry Paris-Saclay,  UMR CNRS  8071,    23 Boulevard de France, 91037 Evry. E-mail: \textcolor{RoyalBlue}{abass.sagna@ensiie.fr} / \textcolor{RoyalBlue}{abass.sagna@univ-evry.fr}.}   
}

\date{}

\begin{document}

\maketitle

\begin{abstract}

The aim of this work is to calculate the fair profit-sharing ratios and the expected payoffs at maturity for each partner in islamic investment contrats (or instruments), based on profit and loss-sharing (PL-sharing). These investment contracts, known as {\em mudarabah} and {\em musharakah}, can  be compared  to {\em limited partnerships} and {\em joint ventures} (including all types of venture, such as  joint-stock companies, partnerships, etc)   in conventional finance. To compute these quantities, we introduce the notion of c-fair profit-sharing ratios, where $c = (c_1, \ldots,c_d) \in (\mathbb R^{\star})^d$  and $d$ is the number of partners. This constitutes an equilibrium approach that accounts for the contributions of the contracting parties in terms of both capital and labour. We show that the $c$-fair profit-sharing ratio of each partner  is the sum of their contributions to capital and labour, weighted by some economic factors that  we identify as {\em investment risk and opportunity}  respectively.    We deduce that, in  the $c$-fair model, the expected investment profit is distributed among the contracts partners  according to the shares  $\varpi_{\ell} = c_{\ell} \, / \, (c_1 +  \ldots + c_d)$, that correspond to their respective contribution weights   to the venture's overall success. We extend these results to mixed contract that combine one or both previous contracts with an agency (known as {\em wakalah}) contract. 

\end{abstract}

%--------------------------------------------------------------------------------------------------------------------
%--------------------------------------------------------------------------------------------------------------------
%---------------------------------OVERVIEW ON OPTIMAL QUANTIZATION--------------------------
%--------------------------------------------------------------------------------------------------------------------
%--------------------------------------------------------------------------------------------------------------------

%that  weight accordingly their payoffs
\section{Introduction}

The goal of this research is to determine the c-fair profit-sharing ratios and expected profit shares of partners in Islamic investment contracts ({\em mudarabah} and {\em musharakah} contracts), possibly combined with a {\em wakalah} (agency) contract. 

In short, A {\em mudarabah} contract is a form of active partnership in which  the capital contributor, or limited partner, provides the capital, while the entrepreneur, or general partner, contributes expertise to help the funds grow. The entrepreneur ({\em mudharib}) contributes to the business venture through his knowledge and skills alone, without investing any property or making   claim to any wages for conducting the business venture (see e.g. \cite{MazSap}).  A {\em musharakah} contract is a partnership between two or more partners who contribute capital to  some profit seeking endeavour.  The two contracts can then   be compared  to {\em limited partnerships} and {\em joint ventures} (including all types of ventures, such as  joint-stock companies, partnerships, etc)   in conventional finance. This broaden the scope of application  of  our work to all forms of joint venture in which the following two principles, common to islamic finance, apply:
 \begin{enumerate} [leftmargin=1.5cm, itemsep=0cm]
 \item[\small{\em  Rule} 1.]   Losses are absorbed by all partners in proportion to their capital contributions.  
  \item[\small{\em  Rule} 2.]  Profit-sharing ratios are determined by mutual agreement between partners.  
 \end{enumerate}
This also  justifies our choice to some times  use  the terms {\em PL-sharing-type  joint venture} and {\em PL-sharing-type  limited partnership}   to refer to {\em musharakah} and {\em mudarabah} contracts, respectively, even through more specific rules of islamic finance regarding these contracts distinguish  them from conventional finance (see e.g. \cite{Askari},  \cite{Ayub},  \cite{Beng}, \cite{Caus} or  \cite{Guer}).

 \begin{remark}{\rm It should be noted  that what distinguishes  a limited partnership  from  a joint venture is capital  contribution. In the joint ventures we are considering, all partners contribute capital, whereas in the limited partnership, the limited partner provides the capital  and the general partner contributes their expertise (or labor) to make it grow. Thus,   {\em Rule } 1. above  states that in the event of a loss in a  {\em PL-sharing-type  limited partnership}, such losses are borne by the limited partner, in exchange for the general partner's loss of the fruits of its labor. This corresponds to the standard framework of islamic finance. 
 
 }
 \end{remark}
 These two financial contracts are the primary sources of the funding in an interest-free economy. However, they are rarely used by financial institutions operating within the framework of islamic finance, and are often replaced by other financial instruments (see, for example, \cite{Dem}, which presents an empirical study of the use of islamic finance instruments by banks in 11 countries). While these alternative instruments are more profitable and (structurally) less risky, their social impact is less significant. Another raison that explains the unpopularity of these funding instruments, and which also explains part of the  associated risk,   is that their economic and financial implications are not well understood. This motivates our work, which goal is to determine the profit-sharing ratios and the expected payoff at maturity for each partner of such a contract. We extend our results to  mixed contracts, which are a combination of one or both of the previous funding instruments and an agency contract. 
 
 The results presented here are new and are bases on a equilibrium approach between the contreparties to the contracts under consideration.  This concept of equilibrium is used in \cite{Lessy} for the specific case of a  {\em PL-sharing mudarabah}, within the framework of a modelling approach that differs from ours (see Remark \ref{Rem-Comparison-Lessy} for a comparison of the two approaches).  In fact, denote by $\gamma_{\ell}$ and by $\kappa_{\ell}$ the profit sharing and the capital contribution ratios, respectively,  of partner  $\ell$. Let $\mbox{Pay}_{\ell}(\gamma_{\ell}, \kappa_{\ell})$ denotes its expected payoff at maturity.  We introduce the notion of $c$-{\em fair} contract, with $c = (c_1, \ldots, c_d)$, $c_{\ell}>0$  for all $\ell$, $d$ standing for the number of partners, that takes into account (making the novelty of the idea with respect to \cite{Lessy}) the contribution of each partner to the project's overall success. The partners' contributions are reflected in a weighted allocation system that assigns a weight coefficient of $1 / c_{\ell}$ to each partner, according to their respective contributions of capital and labour. We then determine the profit-sharing ratios by solving  the equilibrium system $\mbox{Pay}_{1}(\gamma_{1}, \kappa_{1}) \, / \, c_1 \, = \, \ldots \,=\, \mbox{Pay}_{d}(\gamma_{d}, \kappa_{d}) \, / \, c_d$. The profit-sharing ratios resulting from  this equilibrium system are called the  $c$-{\em fair} profit-sharing ratios. 
 
  We show that the $c$-fair profit-sharing ratio of each partner  is the sum of their contributions to capital and labour, weighted by some economic factors that  we identify as {\em investment risk and opportunity}  respectively.  These financial instruments introduce thus  a kind of competition between capital contribution and labor with regard to the investment risk, which increases the share of profit going to capital providers as the risk increases.   We also deduce that, in  the $c$-fair model, the expected investment profit is distributed among the contracts partners  according to the shares  $\varpi_{\ell} = c_{\ell} \, / \, (c_1 +  \ldots + c_d)$, that correspond to their respective contribution weights   to the venture's overall success. We extend these results to mixed contract that combine one or both previous contracts with an agency  contract.

\smallskip

To be more explicit,   consider  a {\em PL-sharing-type joint venture} contract involving $d$ partners. Suppose the partners contributed  amounts   $L_1$, $\ldots$,  and $L_d$, to the capital $L= L_1$ $+ \ldots $ $+ L_d$.  Let  $\kappa_{\ell} = L_{\ell}/L$ be the equity stake of partners $\ell$, for every $\ell \in \{1, \ldots, d\}$ and let  $\gamma_1, \ldots,  \gamma_d$, with $\gamma_1$ $+$ $\cdots$ $+$ $\gamma_d=1$, be the respective profit sharing ratios of the partners.   As previously,  $R_T$ denotes the net income process of the investment  at the maturity $T$ of the contract. Then, the expected  payoff of each partner $\ell$ at the maturity of the contract is given by  
 \begin{equation}  \label{Eq-Payoff-ell-intro}
\mbox{Pay}_{\ell} (\gamma_{\ell},\kappa_{\ell}) =  \gamma_{\ell} \,  \mathbb E(R_T-L)^{+}  - \kappa_{\ell} \, \mathbb E(L-R_T)^{+}.
 \end{equation}
The  fair profit sharing ratios $\gamma_{\ell}$, $\ell=1, \ldots, d$ is determined by solving the system of equations 
\[
\mbox{Pay}_{1} (\gamma_{1},\kappa_{1}) = \ldots = \mbox{Pay}_{\ell} (\gamma_{\ell},\kappa_{\ell}) = \ldots = \mbox{Pay}_{d} (\gamma_{d},\kappa_{d}).
\]
If all the partners contributed equally to the management of the project, we will see that this makes sense and leads to the solution  $\gamma_{\ell} = \kappa_{\ell}$, $\ell=1, \ldots, d$.  However, we know that this is  not usually the case. Therefore,  it is important to understand how to take into account the additional management or labour contributions of certain partners, who may only be compensated for these contributions through their profit-sharing ratios. To illustrate this, let us consider a {\em PL-sharing-type joint venture} contract between two partners whose contributions to the capital are denoted by $\kappa_1$ and $\kappa_2$.  If the partners manage the project  equally, contribution to management, the difference in their overall  contributions to the project lies in their  respective capital contributions, $\kappa_1$ and $\kappa_2$. It is the reasonable to assume that the fair profit-sharing ratios are $\gamma_1 = \kappa_1$ and $\gamma_2 = \kappa_2$. In fact,   in this case, we have
    \begin{eqnarray*}
&& \mbox{Pay}_1 (\gamma_1,\kappa_1)  \, /  \,  \kappa_1   =  \, \mbox{Pay}_2(\gamma_2,\kappa_2) \,  / \,  \kappa_2\quad  \\
 &\mbox{or, equivalently, } &  \mbox{Pay}_1  (\gamma_1,\kappa_1)    =    c  \, \mbox{Pay}_2 (\gamma_2,\kappa_2), \ \mbox{with } c = \kappa_1/ \kappa_2.
   \end{eqnarray*}
   
   Now suppose that the two partners (Partner 1 and Partner 2) contribute the same amount of capital ($\kappa_1=\kappa_2$) but  participate to different degrees in management.  They may then agree to weight  their respective contributions to the project's overall success using the coefficients $c_1$ and $c_2$  to determine the profit-sharing ratios that reflect their respective payoffs. Our aim is therefore  to find $\gamma_1$ and $\gamma_2$ such that  $\mbox{Pay}_{1} (\gamma_{1},\kappa_{1})  \, / \, c_1=  \mbox{Pay}_{2} (\gamma_{2},\kappa_{2})  \, / \, c_2$. We will show   (see Proposition \ref{Prop-Musharakah-d-Partners2Parts}) that, for any $c_1,  c_2>0$,
\begin{eqnarray}
\gamma_{1} &=&   \frac{c_{1}}{c_1+c_2}  \big( 1- \rho(R_T,L) \big) +  \kappa_{1} \,  \rho(R_T,L), \label{EqPart1-1-intro}\\
\gamma_{2} &=&   \frac{c_{2}}{c_1+c_2}  \big( 1- \rho(R_T,L) \big) +  \kappa_{2} \,  \rho(R_T,L), \label{EqPart1-2-intro}
\end{eqnarray}
where $ \rho(R_T,L) =  \mathbb E(L-R_T)^{+}  \, / \, \mathbb E(R_T-L)^{+}$ (observe that $\mathbb E(R_T-L)^{+}>0$ if $R_T \not \equiv L$) is seen  as the {\em investment risk} whereas $1-\rho(R_T,L)$ is seen as the {\em investment opportunity}. It is very important to remark that the {\em viability} assumption   guarantees $\rho(R_T,L) \in [0,1]$ and then, $\gamma_{\ell} \in [0,1]$, for $\ell \in \{1, 2\}$.  Also note that $\gamma_1 + \gamma_2 =1$. 
 \smallskip
 
 Furthermore, setting  $ \varpi_{\ell}   =  c_{\ell}\, / \, (c_1 + c_2)$, $\ell=1, 2$,  we show that  the expected payoff of each partner $\ell \in \{1, 2\}$ is given by
\begin{equation}
\mbox{Pay}_{\ell} (\gamma_{\ell},\kappa_{\ell})  =  \varpi_{\ell} \, \Delta(R_T,L),  \qquad \ell =1, 2.
\end{equation}
In other words,  the $c$-fair profit sharing ratios leads the two partners to  share the  {\em expected investment profit }   $\Delta(R_T,L)$ according the respective weights  $\varpi_{\ell}$, \ $\ell =1, 2$. 

It is also worth noting that a PL-sharing {\em PL-sharing-type  limited partnership} contract appears to be a special case of a {\em PL-sharing-type joint venture} contract where the manager's (or Partner 2's) share of the capital is zero. In fact, setting $\kappa_1 =1 $ and $\kappa_2=0$ in Equations \eqref{EqPart1-1-intro} and    \eqref{EqPart1-2-intro}  i yields the $c$-fair profit-sharing ratios obtained for a {\em PL-sharing-type  limited partnership} contract:
   \begin{eqnarray}
\gamma_{1} &=&   \frac{c_{1}}{c_1+c_2}  \big( 1- \rho(R_T,L) \big) +    \rho(R_T,L), \label{EqPart1-1-intro-Mudhara}\\
\gamma_{2} &=&   \frac{c_{2}}{c_1+c_2}  \big( 1- \rho(R_T,L) \big). \label{EqPart1-2-intro-Mudhara}
\end{eqnarray}
We can compare the previous result with that of  \cite{Lessy} by setting  $c_1 =  c_2$ (see Remark \ref{Rem-Comparison-Lessy} for more details). 
 
 \smallskip
 
 In addition to  applying Equations  \eqref{EqPart1-1-intro} and \eqref{EqPart1-2-intro} to the  {\em PL-sharing-type  limited partnership} contract, it is important to understand the following remarks. Indeed,   for every $\ell \in \{1, 2\}$, we can  decompose it as follows:
\begin{equation}   \label{Eq-Reward-PL-intro}
  \gamma_{\ell} =   \underbrace{ \varpi_{\ell}  \, \big(1-\rho(R_T,L)  \big)}_{\footnotesize \mbox{reward of labour}} + \underbrace{\kappa_{\ell}   \, \rho(R_T,L)}_{\footnotesize \mbox{reward of capital}}.
   \end{equation}
  The previous equation shows that the profit-sharing ratio is :
   \begin{enumerate} [label= $\diamond$]
\item  a function of  investment asset (or investment area), either through investment risk  $\rho(R_T,L)$  or  investment opportunity  $1-\rho(R_T,L)$;
\item  is proportional to the  investment risk $\rho(R_T,L)$  and the invested capital;
\item   is proportional to the  the investment opportunity $1-\rho(R_T,L)$  and  the contribution to labour (or to the management).
\end{enumerate}

From Equation \eqref{Eq-Reward-PL-intro}, we can deduce   that a moderately risky investment (i.e. when the investment opportunity is greater than the investment risk, or equivalently, when $\rho(R_T,L) \le 1/2$) rewards  labour contributions more than  capital contributions. In contrast,  a highly risky investment (i.e. when $\rho(R_T,L) \ge 1/2$) rewards  capital contribution  more. In other words, if a firm is looking to raise capital to finance its business, it is in its best interest to find a highly profitable (or low-risk) activity that will increase its share of the profits.

\smallskip

We also consider the case in which the two partners of a {\em PL-sharing-type joint venture} contract appoint a third partner (Partner 3), who is not one of the original two partners, and in which they are bound to Partner 3 by a {\em PL-sharing-type  limited partnership} contract or a {\em wakalah} contract. In the former case, where management is carried out via a {\em PL-sharing-type  limited partnership}  contract, the $c$-fair profit-sharing ratios are similar to those obtained for a {\em PL-sharing-type joint venture} contract with three partners, with the manager's capital contribution set to zero. When management is arranged via a {\em wakalah}  contract with fixed remuneration (an amount $p$ paid $k$ times up to maturity $T$), if partners 1 and 2 and manager 3 agree to rate their respective payoffs by coefficients $c_1$, $c_2$ and $c_3$, then, the $(c_1,c_2,c_3)$-fair profit sharing  ratios and periodic remuneration $p$ satisfy:
  \begin{eqnarray}  
  \theta(r,T,k) \, p &=&    \varpi_{3}     \, \Delta(R_T,L)\\
 \gamma_{\ell}  &=& \Big( \frac{\varpi_{3} }{2} + \varpi_{\ell}\Big) \,  \big(1-\rho(R_T,L) \big)   + \kappa_{\ell} \, \rho(R_T,L), \qquad  \ell \in \{1, 2\}  \label{EqPart1-Wakala-d-Prop1-intro}  
\end{eqnarray}
where  the coefficient $ \theta(r,T,k)$ comes from the updating value  of the expected payoff of the manager. In fact, we may consider that there is a discounted rate of return $r$ that could be earned from the investment in a good and is given by $ \theta(r,T,k) =  ((1+r )^{ \mbox{\tiny $\frac{T}{k}$} } -1) \, / \, ( (1+r )^{ \mbox{\tiny $T$} } - 1)$.   When there is no discounting, $ \theta(r,T,k) = k$.
%\[
% \theta(r,T,k) =  \frac{(1+r )^{ \mbox{\tiny $\frac{T}{k}$} } -1}{(1+r )^{ \mbox{\tiny $T$} } - 1}, 
%\]
The weights  $(\varpi_{\ell})_{\ell \in \{1, 2, 3 \}}$ still be defined by
\begin{equation}  \label{Eq-weights}
 \varpi_{\ell}= \frac{c_{\ell}}{ c_1 + c_2 + c_3}, \qquad  \mbox{for all } \, \ell \in \{1, 2, 3\}.
\end{equation}
Observe that $\gamma_1 + \gamma_2 =1$.  Furthermore, the expected payoff of each partner $\ell$ (even Partner $3$ who has a fixed $k$-periodic remuneration up to maturity) is given by
\begin{equation}
\mbox{Pay}_{\ell} (\gamma_{\ell},\kappa_{\ell})  = \varpi_{\ell}\, \Delta(R_T,L),  \qquad \ell =1, 2,3.
\end{equation}
 Hence,  the $c$-fair profit sharing ratios leads the three partners to  share the {\em expected investment profit}   $\Delta(R_T,L)$ with the respective weights  $\varpi_{\ell}$,  \ $\ell \in \{1, 2, 3\}$.
 
 A typical application of this result is a situation involving three partners: an investor (Partner 1), who provides all the capital ($\kappa_1=1$); a bank (Partner 2), which acts as the investor's agent through a {\em wakalah} contract, arranging a fixed remuneration  $p$, paid $k$ times up to maturity $T$; and a company (Partner 3), which requires funding and is bound to the bank by a {\em PL-sharing-type  limited partnership} contract ($\kappa_3=0$).  This case is studied in Example \ref{Exam-mudha-wakalah}. 
\smallskip

We also generalize the previous results to contracts involving  multiple partners.  

However, it is important to note that computing the profit-sharing ratios and related quantities in all previous results only requires the computation of the quantities  $\rho(R_T,L)$  and $\Delta(R_T,L)$. These quantities can be  calculated explicitly in certain models, such as when the stochastic process $(R_t)$  is assumed to evolve according to the well-known Black-Scholes model, which assumes, among other things, that the distribution of $R_T$ is log-normal.  Nevertheless, while the values obtained from the Black-Scholes model can be considered as benchmarks, particularly in a lower volatility framework where these values will be close to those expected from more general models, it is well known that most financial datasets are not log-normally distributed. Consequently,  one of the most challenging tasks is to use econometric models, such as ARIMA or (G)ARCH, or more general stochastic diffusion models, such as local volatility or stochastic volatility, to determine the most suitable model for $R$, given an asset dataset. This task will be addressed in a separate paper in order  to reduce the scope of this one.

\medskip

The paper is organised as follows. First, we introduce and discuss the viability assumption and its implications. We then study $c$-fair profit-sharing ratios for a {\em PL-sharing mudarabah} contract and a {\em PL-sharing musharakah} contract. In the latter case, we first consider the matter of two and three partners to help the reader better understand the quantities at stake before generalising.

\section{Viable investment}
To explain what we mean by a "viable investment", consider an investor with initial capital $L$ (which may be in the form of cash or labour). This investor invests in an asset $R$ and expects to receive income $R_T$ at a future time $T > 0$. There are two possible outcomes: 
 \begin{enumerate}
 \item[$\diamond$] The operation succeeds, so the income $R_T$ at time $T$ is  greater than $L$. In this case, the   {\em profit} is   $(R_T - L)^{+}$,
% \begin{equation*}
%  (R_T - L)^{+},
% \end{equation*}
  \item[$\diamond$] The operation  fails, so  $R_T \le L$. In this case, the {\em loss} is   $- (L-R_T)^{+}$.
%  \begin{equation*}
%  - (L-R_T)^{+}.
%   \end{equation*}
 \end{enumerate}
 Let us define the expected profit of such a given  investor as  $ \mathbb E (R_T - L)^{+}$, 
% \begin{equation}
% \mbox{E}_{\mbox{\tiny profit}} (R_T,L) : =  \mathbb E (R_T - L)^{+},
% \end{equation}
 and its expected loss as $  \mathbb E (L - R_T )^{+}$.
% \begin{equation}
% \mbox{E}_{\mbox{\tiny loss}} (R_T,L) : =  \mathbb E (L - R_T )^{+}.
% \end{equation}
% \textcolor{red}{ 

 We also define the  {\em expected investment profit} as  $ \Delta(R_T,L) =  \mathbb E (R_T - L)^{+}   -    \mathbb E (L - R_T )^{+}$, 
% \begin{equation}
% \Delta(R_T,L) =  \mathbb E (R_T - L)^{+}   -    \mathbb E (L - R_T )^{+} ,
% \end{equation}
 {\em the investment risk}  (supposing that $R_T \not \equiv L$, in which case $\mathbb E (R_T - L)^{+}  >0$) as 
 \begin{equation}
 \rho(R_T,L)  =  \frac{   \mathbb E (L - R_T )^{+}  }{  \mathbb E (R_T - L)^{+}   }
 \end{equation}
 and the {\em  expected investment return} (or the {\em investment opportunity}) as $$ \beta(R_T,L):=  \frac{ \Delta(R_T,L)}{  \mathbb E (R_T - L)^{+}}  = 1 - \rho(R_T,L).$$
% \begin{equation}
% \beta(R_T,L) =   \frac{ \Delta(R_T,L) }{  \mathbb E (R_T - L)^{+} }  = 1 - \rho(R_T,L).
% \end{equation}
 %}
 We will say that the investment in asset $R$ with $T$-time maturity is viable if the investor's expected profit over the $T$-term is greater than their expected loss, i.e.  $ \Delta (R_T, L) > 0$. This is equivalent to say that 
\begin{equation}   \label{Def-Rho}
\beta(R_T,L) >  0\quad   \mbox{ or } \quad  \rho(R_T,L)  \in [0,1[. 
\end{equation}
%or, equivalently, 
%\begin{equation}  \label{Def-Rho}
%.
%\end{equation}

It then appears that any $T$-term and $R$-asset-based investment with capital $L$, for which the investment risk   $\rho(R_T,L)$ $ >  1$, is not viable, as its expected loss exceeds the expected profit. Therefore, a $T$-term and $R$-asset based investment (or contract) will be considered to be viable if  statement   \eqref{Def-Rho} holds  true. Consequently, $\rho(R_T.L)$ seems to be inherent in any productive activity, such as trading or manual labour. The quantity $L$ may generally be considered the production costs (or their cash equivalent, as in manual labour, for example). Finally, as previously mentioned, $\rho(R_T.L)$ may also be viewed as the investment risk associated with an activity whose productive costs are $L$ and which generates a random income $R_T$ at time $T$.

\begin{remark} {\rm  We can rewrite $\rho(R_T,L)$ in an universal way as 
\begin{equation}
 \rho(\tilde R_T)  =   \frac{\mathbb E(1-\tilde R_T)^{+} }{ \mathbb E(\tilde R_T-1)^{+}},
\end{equation}
by considering the process $\tilde R = R \, / \,  L$. 
}
\end{remark}

%It then appears that any  $T$-term and $R$-asset based  investment with  capital  $L$ for which  $\rho(R_T,L) >  1$  is not viable because its expected loss is greater than the expected profit. So, a $T$-term and $R$-asset based investment (or contract) will be considered to be viable if the  statement   \eqref{Def-Rho} holds.  Consequently  $\rho(R_T.L)$ seems to be  inherent  to any productive activity like the trading activities or manual labour. The quantity  $L$ may be seen  in general as the production costs (or its equivalence in cash as in a manual labour, for example).  Finally, as  mentioned previously, $\rho(R_T,L)$ may also be seen as the {\em investment  risk} associated to an activity  which productive costs is $L$ and  that generates a random income $R_T$ at time $T$.  

%This motivates the following definition.

 \begin{definition}  A   $T$-term and $R$-asset based investment (or contract) of  capital $L$ is said to be  {\em viable}  if
  \begin{equation}
\Delta (R_T,L) > 0    \quad \mbox{ or, equivalently, if }  \quad  \rho(R_T,L)  \in [0.1[ .
\end{equation}
%The quantity $\rho(R_T,L)$ is called the {\em expected loss-profit ratio} or the {\em investment risk}. 
\end{definition}

We have the following easy result which  characterises   a viable investment (or contract). 

\begin{proposition}   A  $T$-term and $R$-asset based investment (or contract)  of the capital $L$ is   {\em viable} if and only if
\begin{equation}
\mathbb E(R_T) \ge  L.
\end{equation}
\end{proposition}

\begin{proof}[{\bf Proof}.] We have
\begin{eqnarray*}
\Delta (R_T, L)  & = & \mathbb E (R_T - L)^{+}  - \,\mathbb E (L - R_T)^{+} \\
 &=&  \mathbb E \Big((R_T -L) \mathds{1}_{\{  R_T \le L\}} + (L-R_T)\mathds{1}_{\{ R_T > L \}} \Big)  \\
&=&   \mathbb E \Big((R_T -L) \big(1 -  \mathds{1}_{\{  R_T \le L\}} \big)+ (L-R_T)\mathds{1}_{\{ R_T > L \}} \Big)  \\
&=& \mathbb E(R_T -L). 
\end{eqnarray*}
Thus the result.
\end{proof}

\medskip

In the following sections, we will discuss the concept of "$c$-fair" profit-sharing ratios in the context of {\em PL sharing - mudarabah}. To make things easier to understand, we will first study the fair profit-sharing ratios case before introducing the $c$-fair one. First, we will recall the definition of a {\em PL-sharing mudarabah} contract.

%In the next sections, we  discuss the notion of $c$-fair  profit sharing ratios for the {\em PL sharing - mudarabah}. For  ease of understanding, we will study the  fair  profit sharing ratios case before introducing  the $c$-fair one.   We start by recalling the definition of a  {\em PL sharing - mudarabah} contract.
%The results we obtain show in particular  that a  {\em PL sharing - mudarabah} contract may be seen as a {\em PL sharing - musharakah}  contract where the share contribution of the worker to the  capital is null. 

 \section{PL sharing - mudarabah contract}
 We start by describing a {\em mudarabah} contract and then study the  fair profit sharing ratios of these contracts. 
 
 \subsection{What is a {\em mudarabah} contract?}

A {\em mudarabah} contract is a partnership agreement between two entities (which may be, for example, a bank and a company). One partner provides the capital (the owner of which is called the {\em rabb al-maal}), while the other (the {\em mudarib}) provides labour or expertise to manage the funds and generate a profit. Any profit is shared between the partners according to a predetermined key. Each partner may terminate the contract at any time, provided that a fixed maturity date has not been set or the contract has not been executed. Once the contract has been completed and the maturity date has been set, the contract will remain irreversible until the execution of the book value of the cash assets or until the maturity date. However, the profit-sharing ratios may be updated at any time before the maturity date with a mutual agreement between the partners.

%A {\em mudarabah} contract between two entities (which may be, for example, a bank and a company) is a partnership contract where one partner provide the capital (the owner of the capital is called the {\em rabb al-maal}) and the other partner (the {\em mudarib})  brings his labour or  his expertise to manage the funds in order to make a profit. The profit, if any, is shared between the partners  according to an allocation key that must be known at the conclusion of the contract.  Each partner may terminate the contract at any time as long as a maturity is not fixed or before the contract  execution.  After contract completion and  when the maturity is contractual, the contract will still be irreversible up to execution of book value of cash assets or up to the maturity.  However,  with a mutual agreement between the partners,  the profit sharing ratios  may be updated at any time before the maturity.  

 %The profit sharing ratios of the partners must be known at the conclusion of the contract and it may be updated at any time before the maturity  by a mutual agreement between the partners.  
 
 It is important to note that this is a participative contract, since the investor does not contribute labour or expertise, but rather provides capital in exchange for the labour or expertise of the worker.  As a basic principle, the capital owner is only compensated on the profit of the investment, in the same way as the manager. Like the manager (who loses the reward for their labour in case of loss), the capital owner has to support any losses, except in the event of mismanagement (a risk that the owner of the funding may cover by requiring guarantees from the worker).
 
 %Then,  as a basic principle,  the capital owner  is  compensated  only on the profit (in the same way as the manager) of the investment and has, in the same way as the manager who loses the reward of his labour in case of loss,  to support the losses, if any, except in the event of mismanagement (a risk that the funding owner may  cover by requiring guarantees from the worker).  
 %The contract is  then based on truth in the good faith of the worker even if  the funding owner may  cover the risk of mismanagement by requiring guarantees from the worker.   

   Finally, the {\em mudarabah} may be absolute or restrictive. It is restrictive when the funding owner imposes restrictions on the scope of activity or on any other aspect he consider appropriate.  Otherwise, the {\em mudarabah} is said to be absolute.
 
 \smallskip
   
  In the next section, our aim is to determine fair profit-sharing ratios for the partners of a mudarabah contract, in the sense of equalising their payoffs at the maturity of the contract.

 \subsection{Fair profit  sharing ratio} 
 
 Consider a {\em PL-sharing mudarabah} contract between a capital owner (who may be a bank) and a worker (who may be a firm, for example). We denote the profit-sharing ratios between the partners (the funding owner and the worker) by $(\gamma_1, \gamma_2)$, with $\gamma_1 +      \gamma_2=1$. The aim is to determine a fair profit-sharing arrangement between the funding owner and the worker. We denote the net income process of the investment at the maturity $T$ of the contract by $R_T$. Therefore, at the maturity of the contract, 
 \begin{enumerate}
 \item[$\diamond$] the payoff of the funding owner is
 \begin{equation}
 \gamma_1 \,  (R_T-L)^{+}  - (L-R_T)^{+},
 \end{equation}
  \item[$\diamond$] and the payoff of the worker is 
  \begin{equation}
   \gamma_2 \,(R_T-L)^{+}.
   \end{equation}
 \end{enumerate}

 \begin{definition} \label{Def-mudarabah} A $T$-term viable  {\em PL sharing - mudarabah} contract based on asset $R$ is said to be {\em fair} if the funding owner's  expected maturity  payoff  is equal to  the worker's expected payoff. This means that
\begin{equation}  \label{Eq-Def-mudarabah}
 \gamma_1\, \mathbb E[ (R_T-L)^{+}] -  \mathbb E[(L-R_T)^{+}  ]  =  \gamma_2\,\mathbb E[ (R_T-L)^{+} ].
\end{equation}
\end{definition}

Remark that we may also consider the  actualised values of the expected payoffs in Definition \ref{Def-mudarabah} by considering that there is a discounted rate $r$ of return that could be earned from  the investment   on some good of the expected payoffs. However, this is not necessary since Equation \eqref{Eq-Def-mudarabah} will still hold true by simplifying the discounted factor on both sides of the equality.  In the sequel we will consider actualised values of the payoff  when it is necessary. 
\medskip

The following proposition follows directly from  Definition \ref{Def-mudarabah}. 

 \begin{proposition} In a fair {\em PL sharing - mudarabah} contract, the funding owner and the worker  ratios are given by
\begin{eqnarray}  \label{Eq-fair-ratio}
 \gamma_1  & = & \frac{1}{2}  \big(1 +   \rho(R_T,L)\big)\\
 &=&  \frac{1}{2} \big(1  -  \rho(R_T,L)\big)  +   \rho(R_T,L) , \\
 \gamma_2  & = & \frac{1}{2} \big(1  -  \rho(R_T,L)\big),
 \end{eqnarray}
and their common expected payoffs at maturity is given by 
\[
\frac{1}{2} \Delta (R_T,L).
\]
This means that they share equally the {\em expected investment profit }  $ \Delta(R_T,L) =\mathbb E(R_T) - L>0$.
 \end{proposition}
The proof is in fact easy and is given after the following remark, which  discusses on the comparison with  the results of  \cite{Lessy}.

 \begin{remark} \label{Rem-Comparison-Lessy} Define the success probability of the project to be  $\beta = \mathbb P(R_T >L) $. Observing  that $\mathbb E((R_T -L)^{+}) = \beta \, \mathbb E(R_T-L  \vert \{R_T>L\})$,    $\mathbb E((L - R_T )^{+}) = (1-\beta) \, \mathbb E(L - R_T  \vert \{R_T \le L\})$ and setting 
\begin{equation}  \label{Eq-ComparModel-Lessy}
R^{-}(L):=\mathbb E(R_T \vert \{ L \ge R_T \} )  \quad  \mbox{and } \quad  R^{+}(L):=\mathbb E(R_T \vert \{  R_T>L \}),
\end{equation}
  we obtain 
 \begin{equation}  \label{Eq-Mudharah-Compar-Lessy}
 \gamma_{1} =   \frac{1}{2}  \Big( 1+\frac{(1-\beta)}{\beta}\, \frac{L - R^{-}(L)}{ R^{+}(L) -L} \Big).
 \end{equation}
 This must be compared with the result of Proposition 3.2.4. in \cite{Lessy}  which states that if an investor invests an amount $L$ in a company that may generate revenue  $R^{+}(L)>L$ with probability $\beta$, or revenue $R^{-}(L)<L$ with a probability $1-\beta$, then the {\em fair} profit-sharing ratio is 
 \[
 \gamma_1  = \frac{R^{+}(L) - R^{-}(L)}{2 (R^{+}(L) - L) }  +  \frac{R^{-}(L) - L}{2 \beta  (R^{+}(L) - L) } = \frac{1}{2} \Big(  1+ \frac{1-\beta}{\beta}\, \frac{L-R^{-}(L)}{R^{+}(L) - L}  \Big).
 \]
This shows that our model matches with that of  \cite{Lessy} if  the generated revenue $R^{-}(L)$ and $R^{+}(L)$ considered in \cite{Lessy}  are defined as in  \eqref{Eq-ComparModel-Lessy}.
 \end{remark}

\begin{proof}  The proof follows directly from Equation \eqref{Eq-Def-mudarabah}. On the other hand,   replacing in Equation  \eqref{Eq-Def-mudarabah} the ratios $\gamma_1$ and $\gamma_2$ by their values from \eqref{Eq-fair-ratio}, we see that in an $T$-term viable investment of amount $L$ on the asset $R$, the common expected payoff at maturity of both the owner of the capital  and the worker is 
 \[
\frac{1}{2}  \Delta(R_T,L).
 \]
This ends the proof.
\end{proof}

%We may see  from Equation \eqref{Eq-fair-ratio} that in a fair {\em PL  sharing - mudarabah} contract, the profit ratio $\gamma_1$ of the investor is at least of $50\%$, increased by  a quantity that is a non-decreasing function of  the investment risk $\rho(R_T,L)$. 
%  This implies in particular  that in a fair {\em PL sharing - mudarabah} contract, the ratio of the manager  increases with the expected profit of the investment (or decreases with the expected loss). Furthermore,  the larger the expected profit, compared to the expected loss, the bigger will be the ratio of the worker. In the situation where the project is risky such that the expected profit of the investment is quite equal to the expected loss,  $\gamma_1$ will be quite equal to $100\%$, meaning that the manager must accept to work without expecting a paycheck.  This  may happen for a short-term risky, but, (expected) long-term profitable project where the (renewable) {\em mudarabah} contract is structured among different periods for which the  manager would accept a non short-term paycheck.   

  From Equation (20), we can see that, in a fair profit-sharing {\em mudarabah} contract, the investor's profit ratio  $\gamma_1$  is at least $50 \%$, increased by a quantity that is a non-decreasing function of the investment risk $\rho(R_T,L)$. This implies that, in such a contract, the manager's ratio increases with the expected profit  (or decreases with the expected loss) of the investment. Furthermore, the greater the difference between the  expected profit and loss, the greater the worker's ratio will be. If the project is so risky   that the expected profit of the investment is almost equal to the expected loss, then $\gamma_1$ will be almost equal to $100\%$. This  means that the manager must agree to work without pay (for example, during the short term of a high-risky  {\em mudarabah} contract).

%  \medskip
%  
%  \noindent {\em Notation}. From now on, we set 
%  \begin{equation}
%  \rho(R_T) =  \frac{\mbox{  E}_{\mbox{\tiny loss}} (R_T) }{\mbox{E}_{\mbox{\tiny profit}} (R_T) }.
%  \end{equation} 
  \medskip

%  It follows from Equation \eqref{Eq-fair-ratio} that the project is viable for both the funding body and the manager  if the expected loss is strictly less than the expected profit. 
 
% When replacing in Equation  \eqref{Eq-Def-mudarabah} the ratios $\gamma_1$ and $\gamma_2$ by their values from \eqref{Eq-fair-ratio}, we see that in an $T$-term viable investment of amount $L$ on the asset $R$, the common expected payoff at maturity of both the owner of the capital  and the worker is 
% \[
%\frac{1}{2}  \Delta(R_T,L)   ,
% \]
%meaning that they share equally the {\em expected investment profit }  $ \Delta(R_T,L) =\mathbb E(R_T) - L>0$.
%\medskip

Now, consider the situation in which we wish to determine the profit-sharing ratios  $\gamma_1$ and $\gamma_2$ such that the expected payoff  for the capital owner at maturity is $c$ times (with $ c > 0$) the payoff for the worker.  Denote by
\begin{eqnarray*}
&& \mbox{Pay}_{1} (\gamma_1) :=  \gamma_1\, \mathbb E[ (R_T-L)^{+}] -  \mathbb E[(L-R_T)^{+} \\
&\mbox{ and }&  \mbox{Pay}_{2} (\gamma_2) := \gamma_2\, \mathbb E[ (R_T-L)^{+}], 
\end{eqnarray*}
 the respective expected payoffs of the capital owner and the labourer (or the manager). We suppose that
\[
\mbox{Pay}_{1} (\gamma_1)   =   c \,  \mbox{Pay}_{2} (\gamma_2).
\]
Setting $c = c_1/c_2$, this boils down to 
\[
\mbox{Pay}_{1} (\gamma_1) \ / \ c_1  =   \mbox{Pay}_{2} (\gamma_2) \ / \ c_2.
\]

This motivates the introduction of a $(c_1, c_2)$-fair {\em PL sharing - mudarabah} contract, where $c_{\ell}$ is a real number for $\ell = 1, 2$. In the notation of the $(c_1, c_2)$-fair contract, the weighting coefficient$c_{\ell}$  is associated with the payoff $ \mbox{Pay}_{\ell}$. This weighting accounts for each partner's contribution to the project's overall success. As will be seen in Equation \eqref{Eq-Sharing-Weights-mudarabah}, these weighting coefficients determine the weight assigned to each partner in the sharing of the expected common investment profit $\Delta(R_T,L)$.

 \begin{definition} \label{Def-mudarabah-cfair} 
 A $T$-term viable  {\em PL sharing - mudarabah} contract based on  asset $R$  is said to be $(c_1,c_2)$-{\em fair} if \begin{equation}  \label{Eq-Def-mudarabah-c-fair}
 \mbox{Pay}_{1} (\gamma_1)  \ / \ c_1=     \mbox{Pay}_{2} (\gamma_2) \ / \  c_2.
\end{equation}
 \end{definition}

 \begin{proposition}  \label{Prop-Ratios-Profit-Sharing} In a $(c_1,c_2)$-fair {\em PL sharing - mudarabah} contract, the funding owner and the worker profit  ratios are given by
\begin{eqnarray} 
 \gamma_1   &=&  \frac{1}{c_1+c_2}  \big(c_1 + \rho(R_T,L) \, c_2 \big) \nonumber  \\
 & = &   \varpi_1 \,(1 - \rho(R_T,L)  )  +  \rho(R_T,L) ,   \label{Eq-fair-ratio-c-fair1} \\
 \gamma_2  & = & \varpi_2 \, (1 - \rho(R_T,L)  ),   \label{Eq-fair-ratio-c-fair2}
 \end{eqnarray}
 whit $ \varpi_1=c_1 \, / (c_1+c_2)$ and $ \varpi_2=c_2 \, / (c_1+c_2)$. Moreover, their respective expected payoff are given by 
\begin{equation} \label{Eq-Sharing-Weights-mudarabah}
 \frac{c_1}{c_1+c_2}\, \Delta(R_T,L) \quad \mbox{ and  } \qquad   \frac{c_2}{c_1+c_2}\, \Delta(R_T,L).
\end{equation}
This means that they will share the {\em expected investment profit}  $\Delta(R_T,L)$ with the respective weights $ \varpi_1$ and $ \varpi_2$.
 \end{proposition}

\noindent  {\em Comment on Proposition \ref{Prop-Ratios-Profit-Sharing}}. We see from Proposition \ref{Prop-Ratios-Profit-Sharing}  that the weight $\varpi_1$ is the managers's share of the expected investment profit $\Delta(R_T,L)$. It is also the manager's gross compensation for the work when the investment is totally risk-free (i.e. when $\rho(R_T,L) = 0$). Consequence, the weighting coefficients, $c_1$ and $c_2$,  of the overall contributions of the funding owner and the manager to the  project's success can be chosen to meet a desired key of sharing  the expected investment profit $\Delta(R_T,L)$. For example,   the manager's share of the expected profit could be set to be  half that of the funding owner:  $\varpi_1 = 2 \,\varpi_2$, or, equivalently, $c_1 = 2 \, c_2$.

\begin{proof} Keeping in mind that $\gamma_1+\gamma_2=1$,  we  get \eqref{Eq-fair-ratio-c-fair1}-\eqref{Eq-fair-ratio-c-fair2}  by solving Equation \eqref{Eq-Def-mudarabah-c-fair}.  The expected payoff $\mbox{Pay}_{1} (\gamma_1)$ and $\mbox{Pay}_{2} (\gamma_2)$ of the capital owner  and the manager are obtained by replacing $\gamma_1$ and $\gamma_2$ by their values in \eqref{Eq-fair-ratio-c-fair1}-\eqref{Eq-fair-ratio-c-fair2}. 
\end{proof}

\begin{example} In this example we fix  the value of $\rho(R_T,L) \in \{ \frac{1}{4}, \frac{1}{2}\}$ and aim to find the $(c_1,c_2)$-fair profit sharing ratios in a {\em PL sharing - mudarabah} contract between a bank and a firm.  
\smallskip

\noindent 
$\diamond$\, Suppose we want to choose  profit sharing ratios  such that the expected  payoff at maturity is  $\frac{3}{2}$ times greater for the bank  than for the company.  In this case,  we can set $c_1=3$ and $c_2=2$. Furthermore,   the $(c_1,c_2)$-fair profit sharing ratios are given as follows:
\begin{eqnarray*}
& & \mbox{ when }  \ \rho(R_T,L) = \frac{1}{4}, \ \mbox{ then}, \   \gamma_1  = 70\%   \quad  \mbox{ and } \quad \gamma_2  = 30\%,    \\
&\mbox{ and } &  \mbox{ when }  \ \rho(R_T,L) = \frac{1}{2}, \  \mbox{ then}, \  \gamma_1  = 80\%   \quad  \mbox{ and } \quad \gamma_2  = 20\%.  
\end{eqnarray*}
Given the previous selection of ratios $\gamma_1$ and $\gamma_2$, we anticipate that the profit $\Delta(R_T,L)$ will be shared according to the respective weighted shares $(\varpi_1,\varpi_2) = (3 /5, 2/5)$. 
\smallskip

\noindent 
$\diamond$\, Now,  for a fairprofit sharing  (when $(c_1,c_2) = (1,1)$), we get: 
\begin{eqnarray*}
& & \mbox{ when }  \ \rho(R_T,L) = \frac{1}{4}, \ \mbox{ then}, \   \gamma_1  = 62.5\%   \quad  \mbox{ and } \quad \gamma_2  = 37.5\%,    \\
&\mbox{ and } &  \mbox{ when }  \ \rho(R_T,L) = \frac{1}{2}, \  \mbox{ then}, \  \gamma_1  = 75\%   \quad  \mbox{ and } \quad \gamma_2  = 25\%.  
\end{eqnarray*}

This leads to a equal  sharing of the anticipated expected profit  $\Delta(R_T,L)$.
\smallskip

\noindent 
$\diamond$ Finally, we can choose to set the profit-sharing ratios so that the bank can expect to be paid off at maturity $2/3$ times  less than the company expects. In this case, we can set  $c_1=2$ and $c_2=3$, and, we have:
% \,Finally, when we want, because of several extra services developed by the company to ease the management and the tracking of the project for example,  to choose the profit sharing ratios  such that the bank may expect to be paid off at maturity $2/3$ times than  the company may be expecting (we may set in this case  $c_1=2$ and $c_2=3$) then
\begin{eqnarray*}
& & \mbox{ when }  \ \rho(R_T,L) = \frac{1}{4}, \ \mbox{ then}, \   \gamma_1  = 55\%   \quad  \mbox{ and } \quad \gamma_2  = 45\%,    \\
&\mbox{ and } &  \mbox{ when }  \ \rho(R_T,L) = \frac{1}{2}, \  \mbox{ then}, \  \gamma_1  = 70\%   \quad  \mbox{ and } \quad \gamma_2  = 30\%.  
\end{eqnarray*}
Selecting these profit-sharing ratios ($\gamma_1$  for the bank  and $\gamma_2$ for the company) results in the anticipated profit $\Delta(R_T,L)$ being shared according to the respective weighted shares $(\varpi_1,\varpi_2)= (2 /5, 3/5)$. 
\end{example} 

As discussed previously, this example shows that the profit-sharing ratio depends on exposure to risk $\rho(R_T,L)$. Remember that the investor is more exposed to risk than the funding manager because they will have to cover any losses. This exposure to investment risk is reflected in the investor's profit ratio: the riskier the investment, the higher the profit ratio.

%As discussed previously, we can see from this example that the profit-sharing ratio depends on exposure to risk  $\rho(R_T,L)$. Recall that the investor is more exposed to risk than the funding manager because he will have to cover any losses.  This exposure to investment risk is compensated by his profit ratio share: the riskier the investment, the higher the investor profit ratio.

\smallskip

 \section{PL sharing - musharakah contract}
 First, we define a {\em musharakah} contract, then we deal with mixed contracts that combine a {\em musharakah} contract with {\em mudarabah} and {\em wakalah} contracts.
 
 \subsection{Definition}
 
 In a {\em musharakah} contract, two or more partners combine their assets, labour or liabilities in order to generate profits. The share of each partner's capital must be determined in advance. The profit-sharing ratios are agreed by the partners and can be amended before the contract matures. However, all partners absorb losses in proportion to their capital contribution.  A  {\em musharakah}  is said to be {\em diminishing}  if  one partner promises to gradually purchase the other partner's equity share until the title is fully transferred.
 
  In this section, we will determine fair profit-sharing ratios between partners in a {\em PL-sharing musharakah}  contract.
 
% A {\em musharakah} contract involves two or more partners combining their assets, labour or liabilities to generate profits. Each partner's share of the capital must be determined in advance. The profit-sharing ratios are agreed upon by the partners and can be amended at any time before maturity. However, losses are absorbed by all partners in proportion to their capital contribution. In this section, we aim to determine fair profit-sharing ratios between partners in a {\em PL-sharing musharakah} contract.
 
 % \subsection{The $c$-Fair profit  sharing ratios}  
  \subsection{Fair profit sharing ratio of mixed contracts}  
  Suppose there are  two partners who have contributed amounts $L_1$ and $L_2$ to the capital $L$, such that their respective contributions are given by $\kappa_1 = L_1/L$ and $\kappa_2 = L_2/L$.  Let  $\mbox{Pay}_{1} (\gamma_1,\kappa_1)$ and  $\mbox{Pay}_2(\gamma_2,\kappa_2)$  denote the expected payoff of the partners. Then,  
  \begin{eqnarray*}
&&  \mbox{Pay}_1(\gamma_1,\kappa_1)  =  \gamma_{1} \,  \mathbb E (R_T-L)^{+}  - \kappa_{1}\, \mathbb E (L-R_T)^{+} \\
& \mbox{ and } &   \mbox{Pay}_2(\gamma_2,\kappa_2) = \gamma_{2} \,   \mathbb E (R_T-L)^{+}  - \kappa_{2}\,  \mathbb E (L-R_T)^{+},
 \end{eqnarray*}
 where  $\gamma_1$ and  $\gamma_2$ stand  respectively for the profit ratios of partner $1$ and partner $2$. 
 
 We want to determine the  $c$-fair  profit-sharing ratios,  $\gamma_1$ and  $\gamma_2$, and possibly the project manager's remuneration share, depending on the type of contract that binds him to the entrepreneurs.

 First, we consider the situation in which the partners of the {\em PL-sharing musharakah}  contract decide to manage the project themselves.
% We first consider the situation where the  partners of the {\em musharakah} contract  decide to manage the project themselves. 
 
 %Now, let us discuss on the choice of the fair profit sharing ratios $\gamma_1$ and $\gamma_2$ according to the type of management of  the {\em musharakah}  project.
% \begin{enumerate} [leftmargin=*,itemsep=0cm]
\subsubsection{{\em PL-sharing musharakah} contract between two patners}

If the partners are managing the project together and making an equal contribution to management, the difference in their overall contributions lies in their respective capital contributions, $\kappa_1$ and $\kappa_2$. In this case, it would be reasonable to use the following fair profit-sharing ratios:  $\gamma_1 = \kappa_1$ and $\gamma_2 = \kappa_2$. So, we have: 
    \begin{eqnarray*}
&&\mbox{Pay}_1 (\gamma_1,\kappa_1) \ / \  \kappa_1  =   \mbox{Pay}_2(\gamma_2,\kappa_2) \ / \ \kappa_2\quad  \\
 &\mbox{or, equivalently, } &  \mbox{Pay}_1  (\gamma_1,\kappa_1)    =    c  \, \mbox{Pay}_2 (\gamma_2,\kappa_2), \ \mbox{with } c = \kappa_1/ \kappa_2.
   \end{eqnarray*}
   
 Now suppose that the two partners in the {\em PL-sharing musharakah} contract agree to appoint one of them (Partner 2, for example) as manager of the project. In this case, it is not permitted to specify a fixed salary for the manager, who is also a partner in the contract. However, their additional contribution to the management of the project can be rewarded by increasing their profit-sharing ratio. As before, we must determine the $(c_1,c_2)$-fair profit-sharing ratios ($\gamma_1$ and $\gamma_2$)  by selecting the weighting constants  $c_1$ and $c_2$ for their payoffs in a way that offsets Partner 2's additional responsibilities. This means:
%  
%   Now, one question of interest may be to determine the $(c_1, c_2)$-fair profit sharing ratios $\gamma_1$ and $\gamma_2$ heeding the contribution of partner $2$ to the management which is such that one may reasonably consider: 
     \begin{eqnarray*}
&&\mbox{Pay}_1 (\gamma_1,\kappa_1) \ / \  c_1  =   \mbox{Pay}_2(\gamma_2,\kappa_2) \ / \ c_2 \quad  \\
 &\mbox{or, equivalently, } &  \mbox{Pay}_1  (\gamma_1,\kappa_1)    =    c  \, \mbox{Pay}_2 (\gamma_2,\kappa_2), \ \mbox{with } c = c_2/ c_1.
   \end{eqnarray*}
  To determine $\gamma_1$ and $\gamma_2$,  we have to solve the system of equations:
   \begin{equation}  \label{Eq-Example-Musharac1c2}
	   \left\{
	 \begin{array}{lcl}
	 	\mbox{Pay}_1 (\gamma_1,\kappa_1) \ / \  c_1  =   \mbox{Pay}_2(\gamma_2,\kappa_2) \ / \ c_2  \\
		\gamma_1 + \gamma_2 =1, \ \kappa_1 + \kappa_2=1.
	\end{array}
	\right.  
\end{equation}
Its solution reads 
\begin{equation} \label{EqPart1-2}
\gamma_{\ell} = \frac{c_{\ell}}{c_1+c_2} - \left( \frac{c_{\ell}}{c_1 + c_2}  - \kappa_{\ell} \right)\rho(R_T,L), \qquad \ell \in \{1, 2\}.
\end{equation}
%The {\em PL sharing - mudarabah} contract appears then to be a particular case of the  {\em PL sharing - musharakah} contract  where the share contribution of one of the partners (say, partner $2$, considered as the worker) is null. In fact, putting $\kappa_1 =1 $ and $\kappa_2=0$ in Equation \eqref{EqPart1-2} leads to equations \eqref{Eq-fair-ratio-c-fair1}-\eqref{Eq-fair-ratio-c-fair2}. 

Next, we consider the situation in which the two partners of the {\em PL-sharing musharakah} contract appoint an external manager to oversee the project. There are several ways to compensate him, but here we consider two types of contract that could bind the manager to the entrepreneurs: a {\em wakalah} contract, and a {\em mudarabah} contract. 
%   They may  be bound  to the manager by a {\em wakalah} contract

\subsubsection{When the entreprise is managed through a {\em wakalah} contract}
 
 Here, we suppose that the two partners of the {\em PL-sharing musharakah} contract are bound to the manager by a {\em wakalah} contract that provides for fixed remuneration (an amount $p$, paid out $k$ times up to maturity $T$), which will be included in their expenses. In this case, if there is a discounted rate of return $r$ that could be earned from investing in a asset over a $T$-period, the actualised expected payoffs of the partners in the {\em PL-sharing musharakah}  contract, $\mbox{Pay}_{1} (\gamma_1,\kappa_1,p)$ and $\mbox{Pay}_2(\gamma_2,\kappa_2,p)$, satisfy the following, respectively:
   \begin{eqnarray*}
&&  \mbox{Pay}_1 (\gamma_1,\kappa_1,p)  \, = \,  (1+r)^{\mbox{\tiny $-T$}}\, \big(\gamma_{1} \,  \mathbb E (R_T-L)^{+}  - \kappa_{1}\, \mathbb E (L-R_T)^{+} \big)  - \frac{\mbox{Pay}_3 (p,k)}{ 2} \\
 &&  \mbox{Pay}_2 (\gamma_2,\kappa_2,p)  \,  = \,  (1+r)^{\mbox{\tiny $-T$}}\, \big(  \gamma_{2} \,  \mathbb E (R_T-L)^{+}  - \kappa_{2}\, \mathbb E (L-R_T)^{+}\big)  - \frac{\mbox{Pay}_3 (p,k)}{ 2}.
% &&  \qquad \ \  \   \mbox{Pay}_3 (p,k)  \,  = \,  k\,  p. 
   \end{eqnarray*}
In the previous equations,   $\mbox{Pay}_3(p,k)$ is the (actualised expected) payoff of  Partner 3 (which is supposed to be the manager). The quantity  $\mbox{Pay}_3(p,k)$ is  the present value of  the sum of the future cash flow $p$ up to  maturity $T$ of the contract.  So, we have
  \[
  \mbox{Pay}_3(p,k)  = \sum_{i = 1}^k p\, (1+r)^{-{\footnotesize \frac{T}{k}} i }  = p  \, \big(1+r \big)^{ \mbox{\tiny $-T$} }  \, \frac{(1+r )^{ \mbox{\tiny $T$} } - 1  }{(1+r )^{ \mbox{\tiny $\frac{T}{k}$} } -1} \, .
  \]
  As in previous cases, we aim to determine the $(c_1, c_2,c_3)$-fair profit-sharing ratios in a viable {\em PL-sharing musharakah}  contract. In other words, we want to calculate the periodic remuneration $p$ up to maturity and the profit-sharing ratios $\gamma_1$ and  $\gamma_2$ satisfying :
  \begin{equation}  \label{Eq-example-mush-c1c2c3}
    \left\{
	 \begin{array}{lcl}
	 	   \mbox{Pay}_1 (\gamma_1,\kappa_1,p)  \ / \  c_1 =   \mbox{Pay}_2 (\gamma_2,\kappa_2,p) \  /  \ c_2 = \mbox{Pay}_3 (p,k) \ /  \ c_3 \\
		\gamma_1 + \gamma_2 =1.
	\end{array}
	\right.  
  \end{equation}
  If we want  to determine first $p$, it is suitable to rewrite the system \eqref{Eq-example-mush-c1c2c3} as a system of three equations with three unknown: 
  \begin{equation}  \label{Eq-example-mush-c1c2c3-rewrite}
   \left\{
	 \begin{array}{lcl}
	 	\mbox{Pay}_1 (\gamma_1,\kappa_1,p)  \ / \ c_1\, c_2  =   \mbox{Pay}_3 (p,k) \ / \ c_2\,c_3 \\
		    \mbox{Pay}_2 (\gamma_2,\kappa_2,p) \ / \ c_1\, c_2  =   \mbox{Pay}_3 (p,k) \ / \ c_1\,c_3 \\
		\gamma_1 + \gamma_2 =1.
	\end{array}
	\right.  
  \end{equation}
Solving the system \eqref{Eq-example-mush-c1c2c3-rewrite} leads to (setting $  \theta(r,T,k) =  ( (1+r )^{ \mbox{\tiny $\frac{T}{k}$} } -1) \, / \, ( (1+r )^{ \mbox{\tiny $T$} } - 1) $)
\begin{equation}
 \theta(r,T,k)\,  p =    \varpi_3  \, \Delta(R_T,L) , \ \mbox{ with } \quad    \varpi_3 = \frac{c_3}{ c_1 + c_2 +  c_3}.
 % &=& \frac{c_1 c_2}{(c_1 c_2 + c_1 c_3 + c_2 c_3) k}   \,   \big( \mathbb E (R_T)-L\big), 
  \end{equation}
 Then, replacing $p$ by its value in the two first equations gives,  for $\ell \in {\{1,2 \} }$,
%  \begin{equation} 
% % \gamma_{\ell}  = \varpi_{\ell}  - \big (\varpi_{\ell}  - \kappa_{\ell}  \big)  \rho(R_T,L)
%  \gamma_{\ell}  = \Big( \frac{c_{3, \bar 3}}{2} + c_{\ell,\bar{\ell}}\Big) \,  \big(1-\rho(R_T,L) \big)   + \kappa_{\ell} \, \rho(R_T,L), 
%\end{equation}
%  where
%  \begin{equation}  
% \big(c_{1, \bar {1}},\, c_{2, \bar {2}}\big) = \frac{1}{ c_1 c_2 + c_1 c_3 + c_2 c_3} \, \big(c_{2} c_3, \,c_1 c_3 \big).
%\end{equation}
%
 \begin{equation} 
 % \gamma_{\ell}  = c_{\ell, \bar {\ell}}  - \big (c_{\ell, \bar {\ell}}  - \kappa_{\ell}  \big)  \rho(R_T,L)
  \gamma_{\ell}  = \Big( \frac{\varpi_3}{2} + \varpi_{\ell} \Big) \,  \big(1-\rho(R_T,L) \big)   + \kappa_{\ell} \, \rho(R_T,L),
\end{equation}
with
  \begin{equation}  
 \varpi_{\ell} = \frac{c_{\ell}}{ c_1 c_2 + c_1 c_3 + c_2 c_3}.
\end{equation}

%   $\bar{\ell} = \{1,2\} \backslash \{ \ell \}$ and 
%  \[
%  c_{\ell,\bar {\ell}}  =   \frac{c_{ \bar{\ell} }}{c_1 c_2 + c_1 c_3 + c_2 c_3}  \Big( \frac{c_{\ell}}{2}  + c_3\Big).
%  \]
 % \textcolor{red}{with the constraint that $c_{\ell} > 2 c_{3}$, for $\ell \in \{1, 2\}$.}

\subsubsection{When the entreprise is managed through a {\em mudarabah} contract}

Here, we assume that the two partners of the {\em PL-sharing musharakah}  contract are bound by a {PL-profit-sharing  mudarabah} contract with the manager (partner 3).   In this case, the expected payoffs  $\mbox{Pay}_{1} (\gamma_1,\kappa_1)$, $\mbox{Pay}_2(\gamma_2,\kappa_2)$ and $\mbox{Pay}_3(\gamma_3)$ of the partners are given respectively by:
   \begin{eqnarray*}
&& \mbox{Pay}_1 (\gamma_1,\kappa_1)  \, = \,  \gamma_{1} \,  \mathbb E (R_T-L)^{+}  - \kappa_{1}\, \mathbb E (L-R_T)^{+}   \\
 &&  \mbox{Pay}_2 (\gamma_2,\kappa_2)  \,  = \,   \gamma_{2} \,  \mathbb E (R_T-L)^{+}  - \kappa_{2}\, \mathbb E (L-R_T)^{+}   \\
 && \quad \, \  \mbox{Pay}_3 (\gamma_3)  \,  = \,   \gamma_{3} \,  \mathbb E (R_T-L)^{+}. 
   \end{eqnarray*}
 Finding the $(c_1,c_2,c_3)$-fair profit sharing ratio amounts to solve the system
  \begin{equation}  \label{Eq-example-mush-moudharabah-c1c2c3}
    \left\{
	 \begin{array}{lcl}
	 	 \mbox{Pay}_1 (\gamma_1,\kappa_1)  \ / \  c_1    =   \mbox{Pay}_2 (\gamma_2,\kappa_2) \ / \ c_2  =   \mbox{Pay}_1 (\gamma_3,\kappa_3)  \ / \ c_3  \\
		 \hspace{0.32cm}   \gamma_1 + \gamma_2 + \gamma_3=1.
	\end{array}
	\right.  
  \end{equation}
   To compute $\gamma_1$,  we consider the following equivalent system:
    \begin{equation*}  \label{Eq-example-mush-mudaraba-c1c2c3gamma1}
   \left\{
	 \begin{array}{lcl}
	 	 \mbox{Pay}_1 (\gamma_1,\kappa_1)  \ / \  c_1 \,  c_2\, c_3  = \mbox{Pay}_1 (\gamma_1,\kappa_1) \ / \ c_1 \,  c_2\, c_3  \\
		   \mbox{Pay}_1 (\gamma_1,\kappa_1)   \ / \ c_1^2 \,  c_3=   \mbox{Pay}_2 (\gamma_2,\kappa_2) \ / \ c_1 \,  c_2\, c_3 \\
		  \mbox{Pay}_1 (\gamma_1,\kappa_1)   \ / \  c_1^2 \,  c_2  =  \mbox{Pay}_3 (\gamma_3)  \ / c_1 \,  c_2\, c_3 \\
		   \gamma_1 + \gamma_2 +\gamma_3=1.
	\end{array}
	\right.  
  \end{equation*}
 Now, adding the left hand side and then the right hand side of the three first equations of the previous system gives 
 \begin{equation*}
 \Big(\frac{1}{c_2 c_3} + \frac{1}{c_1 c_3} + \frac{1}{c_1c_2} \Big)\, \mbox{Pay}_1 (\gamma_1,\kappa_1)   = \frac{1}{c_2\, c_3} \Big(\mathbb E (R_T-L)^{+}   - \mathbb E (L-R_T)^{+}\Big).
 \end{equation*}
 It follows that
\begin{equation}
     \gamma_1  = \varpi_1  -  \left ( \varpi_1  - \kappa_{1}  \right)  \rho(R_T,L), \qquad \mbox{ with } \  \varpi_1  = \frac{c_1}{c_1 + c_2 + c_3}.
\end{equation}
The ratio $\gamma_2$ is obtained by solving the equivalent system
   \begin{equation*}  \label{Eq-example-mush-mudaraba-c1c2c3gamma1}
   \left\{
	 \begin{array}{lcl}
	 	 \mbox{Pay}_2 (\gamma_2,\kappa_2) \ / \ c_2^2 \,   c_3   =   \mbox{Pay}_1 (\gamma_1,\kappa_1) \ / \ c_1 \,  c_2\, c_3  \\
		  \mbox{Pay}_2 (\gamma_1,\kappa_2) \ / \  c_1 \, c_2 \,  c_3   =  \mbox{Pay}_2 (\gamma_2,\kappa_2)  \ / \ c_1 \,  c_2\, c_3\\
		  \mbox{Pay}_2 (\gamma_2,\kappa_2)   \ / \ c_2^2 \,  c_2  = \mbox{Pay}_3 (\gamma_3)  \ / \ c_1 \,  c_2\, c_3  \\
		   \gamma_1 + \gamma_2 +\gamma_3=1, 
	\end{array}
	\right.  
  \end{equation*}
which solution is given by
\begin{equation}
     \gamma_2  = \frac{c_2}{c_1   + c_2 + c_3 }  -  \left ( \frac{c_2}{c_1   + c_2 + c_3 }  - \kappa_{2}  \right)  \rho(R_T,L).
\end{equation}
Using the condition  $  \gamma_1 + \gamma_2 +\gamma_3=1$ yields 
\begin{equation}
     \gamma_3  = \frac{c_3}{c_1   + c_2 + c_3 }  \Big( 1 -    \rho(R_T,L) \Big).
\end{equation}

% \end{enumerate}
 %\end{enumerate}
 The following proposition summarises all the results of the previous discussions.
 
  \begin{proposition}  \label{Prop-Musharakah-d-Partners2Parts} Consider  a  viable  $T$-term and $R$-asset based  {\em PL-sharing musharakah}   contract between two  partners. Then:
 \begin{enumerate} [leftmargin=*,itemsep=0cm]
 \item If they manage themselves the project  and agree  to rate their respective payoff by the coefficients $1 / c_1$ and $1 / c_2$  according to their contribution to the funding and the management of the project, then, the $(c_1,c_2)$-fair profit sharing ratios are given by 
 \begin{eqnarray}
\gamma_{\ell} &=&  \varpi_{\ell} \,  ( 1- \rho(R_T,L) ) +  \kappa_{\ell} \,  \rho(R_T,L),\\
&=&   \varpi_{\ell}   \, \beta(R_T,L) +  \kappa_{\ell} \,  \rho(R_T,L), \qquad \ell =1, 2,  \label{EqPart1-2-Prop}
\end{eqnarray}
where, for every $\ell \in \{1, 2\}$, \,  $\varpi_{\ell} =  c_{\bar{\ell}} \, / \,  (c_1+c_2).$   Furthermore, the expected payoff of each partner $\ell$  is given by
\begin{equation}
\mbox{Pay}_{\ell} (\gamma_{\ell},\kappa_{\ell})  = \varpi_{\ell}\, \Delta(R_T,L),  \qquad \ell =1, 2.
\end{equation}
\item  Supose they appoint a external manager  and enter into a {\em wakalah} contract with them, whereby the manager receives fixed remuneration (an amount $p$ paid out $k$ times up to maturity $T$). If  Partner 1, Partner 2 and the manager agree to rate their respective payoffs using the coefficients $1/c_1$, $1/c_2$ and $1/c_3$, then the fair profit-sharing ratios and periodic remuneration $p$ will satisfy the following:
\begin{eqnarray}  
  \theta(r,T,k) \, p &=&    \varpi_3   \, \Delta(R_T,L)\\
 \gamma_{\ell}  &=& \Big( \frac{\varpi_3}{2} + \varpi_{\ell} \Big) \,  \big(1-\rho(R_T,L) \big)   + \kappa_{\ell} \, \rho(R_T,L), \,  \ell \in \{1, 2\}. 
\end{eqnarray}
The updating coefficient   $ \theta(r,T,k) =  ((1+r )^{ \mbox{\tiny $\frac{T}{k}$} } -1 ) \, / \, ( (1+r )^{ \mbox{\tiny $T$} } - 1)$ and  the weights  $(\varpi_{\ell})_{\ell \in \{1, 2, 3 \}}$ are given by  $\varpi_{\ell}  =  c_{\ell} \ / \ (c_1 + c_2 + c_3)$, for all $\ell \in \{1, 2, 3\}$.
%\begin{equation}  \label{Eq-Def-c-ell-bar-ell}
% \big(c_{1, \bar {1}},\, c_{2, \bar {2}},\, c_{3, \bar {3}} \big) = \frac{1}{ c_1 c_2 + c_1 c_3 + c_2 c_3} \, \big(c_{2} c_3, \,c_1 c_3, \,c_1 c_2 \big).
%\end{equation}

 Moreover, the expected payoff of each partner $\ell$ (including the manager) is given by
\begin{equation}
\mbox{Pay}_{\ell} (\gamma_{\ell},\kappa_{\ell})  = \varpi_{\ell}\, \Delta(R_T,L),  \qquad \ell \in \{1, 2,3\}.
\end{equation}
\item   If they appoint a manager (Partner  3) outside the company and  enter into  a {\em PL-sharing mudarabah} contract with him, the $(c_1,c_2,c_3)$-fair profit sharing ratios are given  by the following formula (keeping in mind that $\kappa_3=0$)
\begin{equation}
     \gamma_{\ell}  = \varpi_{\ell} \big(1  - \rho(R_T,L) \big)  + \kappa_{\ell}  \, \rho(R_T,L), \quad \mbox{for all } \,\ell \in \{1, 2, 3\},
\end{equation}
with  $\varpi_{\ell} = c_{\ell} \ / \ (c_1 + c_2 + c_3)$.
%\begin{enumerate} [leftmargin=*,itemsep=0cm]
%\item[$\diamond$] for the funding owners (Partner 1 and Partner 2), by
%\begin{equation}
%     \gamma_{\ell}  = \frac{c_3 c_{\bar {\ell}}}{c_2 c_3 + c_1 c_3 + c_1c_2 } \big(1  - \rho(R_T,L) \big)  + \kappa_{\ell}  \, \rho(R_T,L) ,
%\end{equation}
%for  any $\ell \in \{1,2\}$ and where $\bar{\ell} = \{1,2\} \backslash \{ \ell \}$,
%\item[$\diamond$] for the manager, by
%\begin{equation}  \label{EqMush-Mudhar}
%     \gamma_3  = \frac{c_1c_2}{c_2 c_3 + c_1 c_3 + c_1c_2 }  \big( 1 -    \rho(R_T,L) \big).
%\end{equation}
%\end{enumerate}

Still, the expected payoff of each partner $\ell$  is given by
\begin{equation}
\mbox{Pay}_{\ell} (\gamma_{\ell},\kappa_{\ell})  = \varpi_{\ell}\, \Delta(R_T,L),  \qquad \ell \in \{1, 2,3\}.
\end{equation}
%where the $c_{\ell,\bar{\ell}}$'s are defined in \eqref{Eq-Def-c-ell-bar-ell}.
 \end{enumerate}
 \end{proposition}

 \begin{remark}   It is important to note that setting $\kappa_1=1$ and $\kappa_2=0$ in Equation \eqref{EqPart1-2-Prop} yields  the $(c_1,c_2)$-fair profit sharing ratios obtained in Equation \eqref{Eq-fair-ratio-c-fair1} and \eqref{Eq-fair-ratio-c-fair2} for a {\em PL-sharing mudarabah } contract between two partners. Consequently, a  {\em PL-sharing mudarabah} contract between two partners is a {\em PL-sharing musharakah} contract in which one partner provides all the  capital (so that $\kappa_1=1$) and the other provides  labour ($\kappa_2=0$). 
%Remark also that   $\gamma_3$ in Equation \eqref{EqMush-Mudhar} may be written as
%\begin{equation}
%     \gamma_{3}  = \frac{c_1 c_2}{c_2 c_3 + c_1 c_3 + c_1c_2 }  \big(1  - \rho(R_T,L) \big)  +  \kappa_{3}  \,  \rho(R_T,L),
%\end{equation}
%with $\kappa_3=0$. 
Further discussions on the results of  Proposition \ref{Prop-Musharakah-d-Partners2Parts} can be found in the subsequent discussion of Proposition   \ref{Prop-Musharakah-d-PartnersdParts}. 
%We may also seen that in the case where
%Once again, this show that a {\em PL - sharing musharakah } contract with two funding owners partners and a manager (or worker)  bound  to them by a {\em PL-sharing mudarabah} contract is equivalent to a  {\em PL - sharing musharakah } contract with three partners where the share contribution to the capital of one of the partner (whom is the )
 \end{remark}
 
Our current objective is to extend  Proposition \ref{Prop-Musharakah-d-Partners2Parts} to mixed contracts involving multiple partners.  
 
\subsection{Fair profit  sharing ratio: the general framework}

 We consider a {\em PL-sharing musharakah} contrat between $d$  partners: partner  1,  Partner 2, $\ldots$,   and  partner $d$. When $d=2$, the partnership  may be for example  between a bank and a company,  two companies,  two banks, etc. The partners  contributed respectively amounts    $L_1$, $L_2$, $\ldots$,  and $L_d$,  to the  capital.  Let $L=L_1+ \cdots + L_d$ be the initial  capital and define  $\kappa_{\ell} = L_{\ell}/L$  as the equity stake of partners $\ell$, for every $\ell \in \{1, \ldots, d\}$. Finally,   let us denote the profit sharing ratios of the partners by  $(\gamma_1, \ldots,  \gamma_d)$, with $\gamma_1$ $+$ $\cdots$ $+$ $\gamma_d=1$. The aim is  to determine the fair profit sharing between the  partners.  
 
 \medskip
 
 As before,  $R_T$ denotes the net income process of the investment  at  maturity $T$.  At the maturity of the contract, the expected payoff of partner $\ell$ is given by 
 \begin{equation}  \label{Eq-Payoff-ell}
\mbox{Pay}_{\ell} (\gamma_{\ell},\kappa_{\ell}) =  \gamma_{\ell} \,  \mathbb E(R_T-L)^{+}  - \kappa_{\ell} \, \mathbb E(L-R_T)^{+}.
 \end{equation}
%  \item[$\diamond$] the payoff of Partner $2$  is 
% \begin{equation}
% \gamma_2 \,  (R_T-L)^{+}  - \kappa_2 (L-R_T)^{+}.
% \end{equation}
% \end{enumerate}
We next  define a $(c_1, \ldots, c_d)$-fair {\em PL-sharing musharakah}  contract. 
 
 \begin{definition} \label{Def-musharakah} A  viable $T$-term and $R$-asset based  {\em PL-sharing musharakah} contract involving  $d$ partners is said to be $(c_1, \ldots,c_d)$-{\em fair} if 
\begin{equation}  \label{Eq-Def-musharakah}
  \mbox{Pay}_1 (\gamma_1,\kappa_1)  \ / \ c_1 \ = \ \cdots \ = \   \mbox{Pay}_d (\gamma_d,\kappa_d) \ /   c_{d} .
\end{equation}
 \end{definition}

%Now, define some sets that will be used in the sequel: we denote the set 
%\begin{equation*}
%\mathcal{P}_d = \{1, \ldots, d\}     \qquad \mbox{ and } \quad  \bar{\mathcal {P}}_d (\ell) = \mathcal {P}_d \backslash \{\ell\}.
%\end{equation*}

The following result gives the profit sharing ratio of each partner of a  {\em PL-sharing musharakah} contract. We still define the weights  $(\varpi_{\ell})_{1 \le \ell \le d}$  by
\begin{equation}  \label{Eq-weights}
\varpi_{\ell}  = c_{\ell} \   \big /  \, \sum_{i=1}^d c_i ,  \,  \mbox{for every } \  \ell \in \{1, \ldots, d\}.
\end{equation}

  \begin{proposition}  \label{Prop-Musharakah-d-PartnersdParts} Consider  a  viable   $T$-term and $R$-asset based  {\em PL-sharing musharakah}  contract between  $d$ partners whom manage the project  themselves   and agree   to rate their respective payoff by the coefficients $1/ c_{\ell}$, $\ell \in \{1, \ldots, d\}$. Then, the $(c_1,\ldots, c_d)$-fair profit sharing ratios are given by
 \begin{eqnarray} 
 \gamma_{\ell}  &=& \varpi_{\ell}  \,\beta(R_T,L) + \kappa_{\ell}   \, \rho(R_T,L) \label{EqPart1-d-Prop1}  \\
  &=& \varpi_{\ell}  \,\big(1-\rho(R_T,L) \big) + \kappa_{\ell}   \, \rho(R_T,L),   \quad \mbox{ for any } \ \ell \in  \{1, \ldots, d\}. \label{EqPart1-d-Prop2}
\end{eqnarray}
Furthermore, the expected payoff of each partner $\ell$ is given by
\begin{equation}
\mbox{Pay}_{\ell} (\gamma_{\ell},\kappa_{\ell})  = \varpi_{\ell}\, \Delta(R_T,L),  \qquad \ell =1, \ldots,d.
\end{equation}
This means that the partners will share the {\em expected investment profit}   $\Delta(R_T,L)$ according to the respective weights  $\varpi_{\ell}$.
 \end{proposition}
\medskip

%One interesting case is when $d-1$ partners of a PL-sharing musharakah contract appoint a manager (partner $d$) who is not one of them, and are bound to him by a PL-sharing mudarabah contract. This appears to be a particular case of Proposition  \ref{Prop-Musharakah-d-PartnersdParts}, which we state as a corollary.

An interesting case arises when $d-1$ partners in a {\em PL-sharing musharakah}  contract appoint a manager (partner $d$), who is not one of the partners, and enter into a {\em PL-sharing mudarabah} contract with him. This appears to be a special instance of Proposition \ref{Prop-Musharakah-d-PartnersdParts}  which we present as a corollary.

  \begin{corollary}  \label{Cor-Musharakah-d-PartnersdParts} Consider  a  viable   $T$-term and $R$-asset based  {\em PL-sharing musharakah}  contract involving  $d-1$ partners. Suppose the partners appoint a manager (partner $d$) from outside their group and enter into  a {\em PL-sharing mudarabah} contract.  If they agree to weight their respective payoff by the coefficients $1/ c_{\ell}$, $\ell \in \{1, \ldots, d\}$, according to their contribution to the funding and the management of the project,  the $(c_1,\ldots, c_d)$-fair profit sharing ratios are given by
 \begin{eqnarray} 
&&  \gamma_{\ell} \  = \varpi_{\ell}  \,\big(1-\rho(R_T,L) \big) + \kappa_{\ell}   \, \rho(R_T,L),  \quad \ell=1, \ldots,d-1,  \label{EqPart1-d-Cor1}   \\
  &&   \gamma_{d} \  = \ \varpi_{d}   \,\big(1-\rho(R_T,L) \big).    \label{EqPart1-d-Cor2}
\end{eqnarray}
Furthermore, the expected payoff of each partner $\ell$ is given by
\begin{equation}
\mbox{Pay}_{\ell} (\gamma_{\ell},\kappa_{\ell})  = \varpi_{\ell}\, \Delta(R_T,L),  \qquad \ell =1, \ldots,d.
\end{equation}
 \end{corollary}
\medskip

Before giving the proof of Proposition \ref{Prop-Musharakah-d-PartnersdParts}, let us make the following remarks.  
\medskip

 \begin{remark}  \label{Rem-MainResult-1} 
%Recall that  the coefficients $(c_{\ell})_{\ell \in \mathcal{P}_d}$ are put to rate  the contribution on the management of the project.  The associated coefficients we define as the function 
% \begin{eqnarray*}
% &&  \mbox{w} : \mathcal{P}_d \longrightarrow \mathcal{P}_d\\
% &&                    \hspace{0.55cm} c_{\ell}\, \longmapsto \mbox{w} (c_{\ell}) =   \varpi_{\ell}
% \end{eqnarray*}
%  are the corresponding sharing weights  of the {\em expected investment profit}  $\Delta(R_T,L)$  (note in fact that $\mbox{w} (c_{1}) + \cdots + \mbox{w} (c_{d}) =1$).  It is clear from Equation \eqref{Eq-weights} that the smaller is the rate coefficient $c_{\ell}$, the bigger will be the associated weight $\mbox{w} (c_{\ell}) =   \varpi_{\ell}$.  
%  

 Owing to Equation \eqref{EqPart1-d-Prop2} (or  to equations \eqref{EqPart1-d-Cor1} and \eqref{EqPart1-d-Cor2}), we can see that the $(c_1, \ldots,c_d)$-fair profit-sharing ratio $\gamma_{\ell}$ of  partner $\ell$ can  be decomposed as
  \begin{equation*}
  \gamma_{\ell} =   \underbrace{ \varpi_{\ell}  \, \big(1-\rho(R_T,L)  \big)}_{\footnotesize \mbox{reward fof  labour}} + \underbrace{\kappa_{\ell}   \, \rho(R_T,L)}_{\footnotesize \mbox{reward of capital}}.
   \end{equation*}
Therefore,  tthe $(c_1, \ldots,c_d)$-fair profit-sharing ratio  appears to be a reward for the contribution to  capital  ($\kappa_{\ell}   \, \rho(R_T,L)$)   and a reward for  the contribution to the management (or labour) of the project  ($\varpi_{\ell}  \,(1-\rho(R_T,L) )$). We can also deduce that a moderately risky investment (i.e. an investment opportunity greater than the investment risk, or equivalently, $\rho(R_T,L) \le 1/2$) rewards project management contributions  more than capital participation, whereas a high-risk investment (i.e. $\rho(R_T,L) \ge 1/2$) rewards  capital contributions more. In other words, if we (a firm, for example) are looking to raise capital to finance our business, it is in our best interests to find a highly profitable (or low-risk) activity that will increase our share of the profits.
 \end{remark}
 \begin{corollary}  \label{Cor-Compar-Gamma}
We still deduce from Equation  \eqref{EqPart1-d-Prop2}  that  for any $\ell,\ell' \in \{1, \ldots, d  \}$,
\begin{equation}
\gamma_{\ell} - \gamma_{\ell'}  = (\varpi_{\ell} - \varpi_{\ell'}) \,\big(1-\rho(R_T,L) \big) +  (\kappa_{\ell} - \kappa_{\ell'} ) \,\rho(R_T,L).    \label{EqPart1-d-Prop-Difference-Gamma}
\end{equation} 

 Then, the difference in profit-sharing ratios between partners $\ell$ and $\ell'$ ($\gamma_{\ell} - \gamma_{\ell'} $)  is  balanced by  the difference in their contributions to capital ($\kappa_{\ell} - \kappa_{\ell'}$)  and the difference in their  weighted  contributions to  management ($\varpi_{\ell} - \varpi_{\ell'}$). Consequently, we deduce from Equation \eqref{EqPart1-d-Prop-Difference-Gamma} that:
\begin{enumerate}[leftmargin=*,itemsep=0cm]
\item If $\kappa_{\ell}  \ge \kappa_{\ell'}$ and $\varpi_{\ell}  \ge \varpi_{\ell'}$ (i.e., the partner $\ell$ contributes  more than partner $\ell'$ in terms of  both  capital and  labour) then, $\gamma_{\ell} \ge \gamma_{\ell'}$.
\item  If  $\kappa_{\ell}  <  \kappa_{\ell'}$ and  $\varpi_{\ell}   >  \varpi_{\ell'} $, that is, partner $\ell$ contributes less  capital  than partner $\ell'$   but, more labour to the project's overall success, then,  
% $\kappa_{\ell}  -  \kappa_{\ell'} <  c_{\ell, \bar {\ell}}  - c_{\ell', \bar {\ell'}}$, then,   
\begin{equation} \tag{9}
\gamma_{\ell}  \ge \gamma_{\ell'}  \qquad \Longleftrightarrow \qquad \rho(R_T,L)\, \le \,   \frac{\varpi_{\ell}  - \varpi_{\ell'}}{ (\varpi_{\ell}  - \varpi_{\ell'})  - (\kappa_{\ell} - \kappa_{\ell'})} \in (0,1).
\end{equation}
If in addition,  $\rho(R_T,L) \le   1 / 2$,   $\varpi_{\ell}  -  \varpi_{\ell'}  > \kappa_{\ell'}  -  \kappa_{\ell}$  (partner  $\ell$ bridges the funding gap for capital through his management contributions), then, $\gamma_{\ell}  \ge \gamma_{\ell'}.$
%\item  If $\kappa_{\ell}  \le \kappa_{\ell'}$, then, $\gamma_{\ell}  \ge \gamma_{\ell'}$  as soon as $c_{\ell, \bar {\ell}}  > c_{\ell', \bar {\ell'}}$ and 
%\begin{equation*}
%\rho(R_T,L)\, \le \,   \frac{c_{\ell, \bar {\ell}}  - c_{\ell', \bar {\ell'}}}{ (c_{\ell, \bar {\ell}}  - c_{\ell', \bar {\ell'}})  + (\kappa_{\ell'} - \kappa_{\ell})}.
%\end{equation*}
\item If  $\varpi_{\ell}  <  \varpi_{\ell'}$  and $\kappa_{\ell}  >  \kappa_{\ell'}$  (partner  $\ell$ contribues less in terms of  management  than partner $\ell'$, but more  in terms of capital), then, 
\begin{equation} \tag{10}
\gamma_{\ell}  \ge \gamma_{\ell'}  \qquad \Longleftrightarrow \qquad \rho(R_T,L)\, \ge \,   \frac{\varpi_{\ell}  - \varpi_{\ell'}}{ (\varpi_{\ell}  - \varpi_{\ell'})  - (\kappa_{\ell} - \kappa_{\ell'})} \in (0,1).
\end{equation}
If furthermore,  $\rho(R_T,L)\, \ge  \,  1/2$,  $\kappa_{\ell}  -  \kappa_{\ell'} >  \varpi_{\ell'} - \varpi_{\ell}$  (partner  $\ell$  closes the gap in management contributions through his capital contribution), then, $\gamma_{\ell}  \ge \gamma_{\ell'}$.
\end{enumerate}

 \end{corollary}

%Then, when two different partners $\ell$ and $\ell'$ have the same share contributed to the  capital so that $\kappa_{\ell} = \kappa_{\ell'}$ (or $\kappa_{\ell} \ge \kappa_{\ell'}$), the one that has contributed more to the management of the project  will have the highest profit ratio. In fact, if the partner $\ell$ has bigger contribution to the management of the project than partner $\ell'$, means,  $c_{\ell} < c_{\ell'}$, it follows from   Equation \eqref{Eq-weights} that $\varpi_{\ell} > c_{\ell', \bar {\ell'}}$. As a consequence, $\gamma_{\ell}>\gamma_{\ell'}$ (as it will appear in Example \ref{Example-Musharaka}). We also see that when $\kappa_{\ell} = \kappa_{\ell'}$ (or $\kappa_{\ell} \ge \kappa_{\ell'}$), the gap $\gamma_{\ell} - \gamma_{\ell'} $ between the profit ratios of partners $\ell$ and $\ell'$ decreases with the  investment risk $\rho(R_T,L)$. The more the investment is risky, the lower will be the supra profit ratio of partner $\ell$ with respect partner $\ell'$.  This is because  the     he exposes his labour or expertise contribution to the management to a loss when the investment is risky.  We see here that . 
%
%This is because if $\kappa_{\ell} \ge \kappa_{\ell'}$, up to a given level of contribution to the management, the investment risk still be of much borne by the  partners that contributed the more to the funding of the capital.   he exposes his labour or expertise contribution to the management to a loss when the investment is risky.  We see here that . 

% \end{remark}

 \begin{remark}  \label{Rem-MainResult-3}
%\medskip

We may also note that when   $c_{i} =1$, for any $\ell \in \{1, \ldots, \}$,  we have  $\varpi_{\ell}  = 1/d$. Equation \eqref{EqPart1-d-Prop2} becomes 
\begin{equation} 
\gamma_{\ell}  = \frac{1}{d}  - \Big(\frac{1}{d} - \kappa_{\ell } \Big) \rho(R_T,L).
\end{equation}
In this case, we have  for any   $\ell, \ell' \in \{1, \ldots, d\}$, \ $\gamma_{\ell} - \gamma_{\ell'}  = (\kappa_{\ell} - \kappa_{\ell'}) \rho(R_T,L),$
%\begin{equation}
%\gamma_{\ell} - \gamma_{\ell'}  = (\kappa_{\ell} - \kappa_{\ell'}) \rho(R_T,L),
%\end{equation}
% Furthermore, since  for any $\ell \not= \ell'$,
%\[
%\gamma_{\ell}  - \gamma_{\ell'} = (\kappa_{\ell } - \kappa_{\ell'})\, \rho(R_T,L),
%\]
so that  the difference in the profit-sharing ratio between two partners comes from the difference in their share of the capital contribution. If all partners  have contributed the same amount of   capital  ($\kappa_{\ell} = \kappa_{\ell'}$, for any $\ell \not= \ell'$), then the fair profit-sharing ratio will be equal for all partners, regardless of the value of  $\rho(R_T,L)$: $\gamma_{\ell} = \frac{1}{d}$, for any  $\ell \in \{1, \ldots, d\}.$  This also suggests that the best way to absorb investment risk in a {\em PL-musharakah} contract is to encourage equal participation in capital and labour. 
%\[
%\gamma_{\ell} = \frac{1}{d}, \quad  \mbox{ for any } \ell \in \{1, \ldots, d\}.
%\]
%This means in particular that the best way to manage (or, the only way to neutralise 1) the investment risk
  \end{remark}

  \begin{figure}[htpb]
 \begin{center}
  \!\includegraphics[width=11cm,height=7cm]{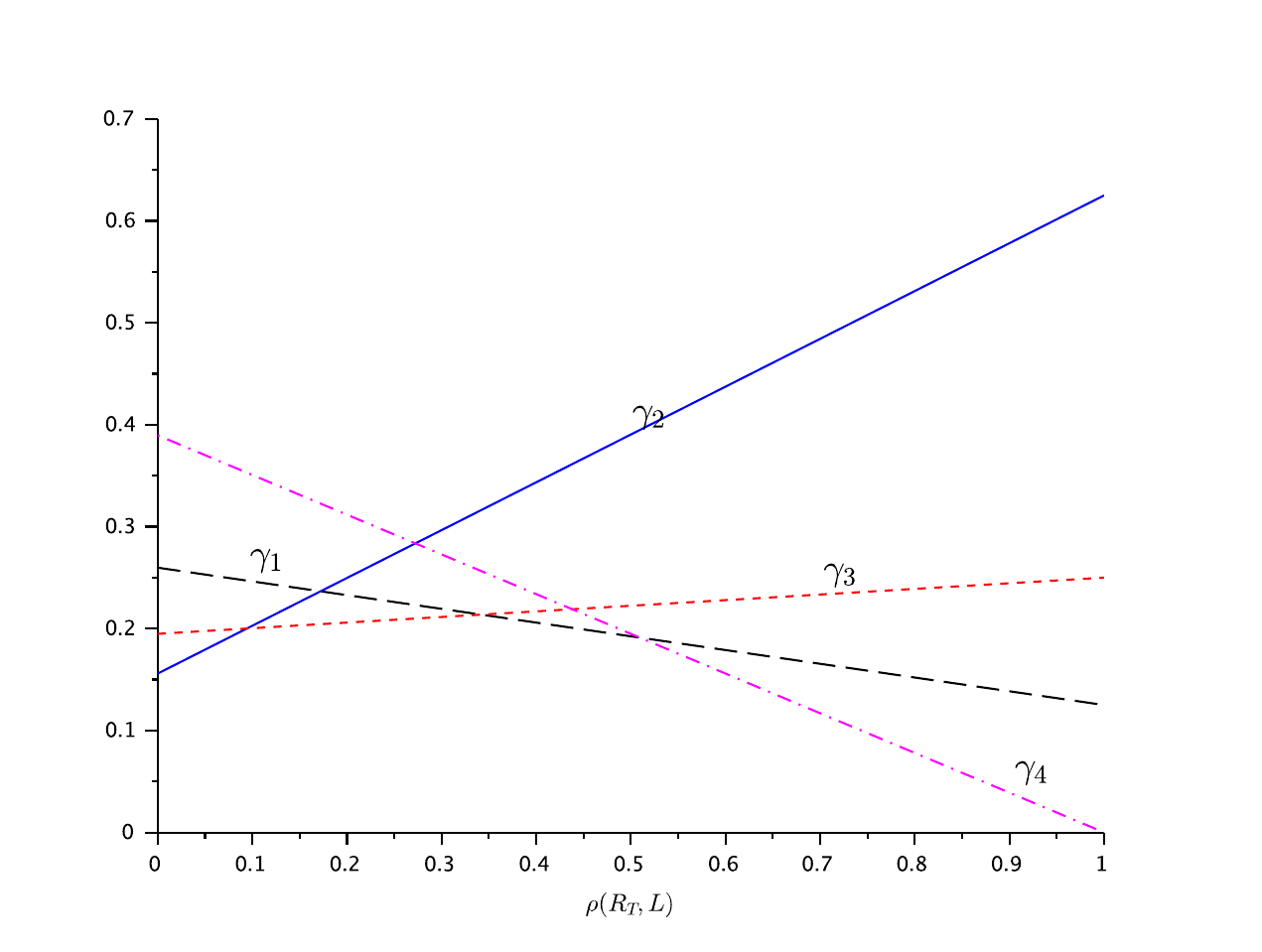}	
%\hfill   \!\includegraphics[width=6.0cm,height=6.0cm]{ProcPoissLambda2.pdf}
  \caption{\footnotesize  We represent the $(c_1,c_2,c_3,c_4)$-fair profit sharing ratios $\gamma_{\ell}$ (in ordinate axis) as functions of the investment risk  $\rho(R_T,L)$ (in abscissa axis):  $(c_1,c_2,c_3,c_4) = (3,5,4,2)$ and $(\kappa_1, \kappa_2,\kappa_3,\kappa_4) = (1/8,5/8,2/8,0)$.}
  \label{fig-PL-Sharing}
  \end{center}
\end{figure}

Figure \ref{fig-PL-Sharing}  depicts the $(c_1,c_2,c_3,c_4)$-fair profit sharing ratio as a function of the investment risk $\rho(R_T,L)$. We set  $(c_1,c_2,c_3,c_4) = (3,5,4,2)$ and $(\kappa_1,$ $ \kappa_2,$ $\kappa_3,$ $\kappa_4)$ $ = (1/8,$ $5/8,$ $ 2/8,$ $0)$. Depending on the values taken by the $c_{\ell}$'s and the $\kappa_{\ell}$'s, we  see that the profit sharing ratios may  be  either  non-increasing or  non-decreasing functions of $\rho(R_T,L)$.  This is in line with  Corrolary  \ref{Cor-Compar-Gamma}.  

% \end{remark}
\medskip

Now, let us prove Proposition \ref{Prop-Musharakah-d-PartnersdParts}.

\begin{proof}  (of Proposition \ref{Prop-Musharakah-d-PartnersdParts})] Let $\ell \in \{1, \ldots, d\}$. Since  the contract is $(c_1, 
\ldots, c_d)$-fair, we have in particular
   \begin{equation*} 
   \left\{
	 \begin{array}{lcl}
	 	  \mbox{Pay}_{\ell} (\gamma_{\ell},\kappa_{\ell}) \ / \  c_{\ell}   =   \mbox{Pay}_1 (\gamma_1,\kappa_1)  \ /  \ c_1   \\
		 \mbox{Pay}_{\ell} (\gamma_{\ell},\kappa_{\ell})   \ / \  c_{\ell}  =  \mbox{Pay}_2 (\gamma_2,\kappa_2) \ / c_2   \\
		\quad \vdots   \qquad \vdots \qquad \vdots  \qquad \quad  \vdots   \qquad \vdots  \\    
		   \mbox{Pay}_{\ell} (\gamma_{\ell},\kappa_{\ell})   \ / \  c_{\ell}   =  \mbox{Pay}_d (\gamma_d,\kappa_d) \ / c_d  \\
		   \gamma_1 + \ldots +\gamma_d=1.
	\end{array}
	\right.  
  \end{equation*}
Il follows that
  \begin{equation*} 
   \left\{
	 \begin{array}{lcl} 
	 	\mbox{Pay}_{\ell} (\gamma_{\ell},\kappa_{\ell})  \ / \ \big(c_{\ell} \, \prod_{j \not=1} c_j \big)  =   \mbox{Pay}_1 (\gamma_1,\kappa_1) \ / \ \big( \prod_{j =1}^d c_j \big)   \\
		\mbox{Pay}_{\ell} (\gamma_{\ell},\kappa_{\ell})  \ / \ \big(c_{\ell} \, \prod_{j \not=2} c_j \big)  =    \mbox{Pay}_2 (\gamma_2,\kappa_2)  \ / \ \big( \prod_{j =1}^d c_j \big) \\
		\quad \vdots   \qquad \vdots \qquad \vdots  \qquad \quad  \vdots   \qquad \vdots  \\    
		\mbox{Pay}_{\ell} (\gamma_{\ell},\kappa_{\ell})  \ / \ \big(c_{\ell} \, \prod_{j \not=d} c_j \big)    =     \mbox{Pay}_d (\gamma_d,\kappa_d) \ / \ \big( \prod_{j =1}^d c_j \big) \\
		   \gamma_1 + \ldots +\gamma_d=1.
	\end{array}
	\right.  
  \end{equation*}
Adding the left and right hand sides of the $d$ first equations of the previous system gives (setting $E_1: =  \mathbb E(R_T-L)^{+}$ and $E_2:=\mathbb E(L-R_T)^{+}$)
\begin{equation*}
 \big( \gamma_{\ell} \, E_1   - \kappa_{\ell}\, E_2  \big)  \  \frac{1}{ c_{\ell}}  \Bigg(\sum_{i=1}^d  \frac{1}{\prod_{j \not=i} c_j} \Bigg) =\big( E_1  -   E_2 \big) \ / \  \Big(\prod_{i=1}^d c_i \Big).
\end{equation*}
We deduce that 
\begin{eqnarray*}
 \big( \gamma_{\ell} \, E_1   - \kappa_{\ell}\, E_2  \big)  \  \frac{1}{ c_{\ell}}  \Bigg(\sum_{i=1}^d  \frac{c_i}{\prod_{j =1}^d c_j} \Bigg) =\big( E_1  -   E_2 \big) \ / \  \Big(\prod_{i=1}^d c_i \Big).
\end{eqnarray*}
Consequently,
\begin{eqnarray*}
 \big( \gamma_{\ell} \, E_1   - \kappa_{\ell}\, E_2  \big)  \  \frac{1}{ c_{\ell}}  \sum_{i=1}^d  c_i =\big( E_1  -   E_2 \big).
\end{eqnarray*}
Therefore,  \, $ \gamma_{\ell} \, E_1   - \kappa_{\ell}\, E_2   = \varpi_{\ell}\big( E_1  -   E_2 \big),$
and  this leads to Equation \eqref{EqPart1-d-Prop2}. 

Now, the expected payoff $\mbox{Pay}_{\ell} (\gamma_{\ell},\kappa_{\ell})$ of each partner $\ell$ is obtained by replacing in Equation \eqref{Eq-Payoff-ell},   $\gamma_{\ell}$ by its value in \eqref{EqPart1-d-Prop2}.
\end{proof}

\begin{example} {\rm  \label{Example-Musharaka1}  Consider a viable {\em PL-sharing musharakah} contract of four partners,  with time horizon $T$, based on the asset $R$. Suppose the partners agree to weight their contrubution to the venture's success   by the  coefficients $1/c_{\ell}$, with $\ell=1, \ldots, 4$. We determine $c$-fair profit sharing ratios  for given values of $c_{\ell}$, when  $\rho(R_T) \in \{\frac{1}{8},  \frac{3}{4}\}$.

\medskip

% Suppose  all the partners contributed to the same amount of  capital: $\kappa_{\ell} = 25\%$, for any $\ell \in \{1, 2, 3, 4\}$. 
\begin{enumerate} [leftmargin=*,itemsep=0cm]
\item Assumung that all partners contribute  the same share  of $\kappa_{\ell} = 25\%$ of the capital, we have: 
\begin{enumerate} [leftmargin=0.3cm,itemsep=0cm]
\item[$\diamond$] If  $(c_1, c_2,c_3,c_4)=(1,1,1,1)$, then $(\varpi_{1},$ $ \varpi_{2},$ $\varpi_{3},$ $\varpi_{4})=$ $(1/4, $ $1/4, $ $1/4, $ $1/4)$ and $\gamma_{\ell} = 25\%, \, \ell \in \{1, \dots, 4\}$,  . 
%Then,   the expected investment profit allocation key  equals $(\varpi_{1},$ $ \varpi_{2},$ $\varpi_{3},$ $\varpi_{4})=$ $(1/4, $ $1/4, $ $1/4, $ $1/4)$. Regardless of the value taken by $\rho(R_T,L)$, the $(c_1, \ldots,c_4)$-fair profit sharing ratios are equal (as discussed in Remark \ref{Rem-MainResult-3}): $\gamma_{\ell} = 25\%, \, \ell \in \{1, \dots, 4\}$.
%\[
%\gamma_{\ell} = 25\%, \qquad \mbox{for any partner } \ell \in \{1, \dots, 4\}.
%\] 
\item[$\diamond$] If  $(c_1, c_2,c_3,c_4)=(4,2,4,1)$,  then  $(\varpi_{1},$ $ \varpi_{2},$ $\varpi_{3},$ $\varpi_{4})=$ $\frac{1}{11} \, (4, $ $2, $ $4, $ $1)$ and
\begin{eqnarray*}
&& (\gamma_1, \gamma_2, \gamma_3, \gamma_4)  \ = \  (35\%,\ 19 \%,\ 35\%, \ 11\%) \qquad \mbox{ if } \, \rho(R_T) =1/8 \\
&\mbox{ and }& (\gamma_1, \gamma_2, \gamma_3, \gamma_4)  \ \simeq  \  (29\%,\ 20 \%,\ 29\%, \ 22 \%) \qquad \mbox{ if } \, \rho(R_T) =2/3. 
\end{eqnarray*}
Since $ \kappa_{\ell} = \kappa_{\ell'}$, for all $\ell \not= \ell'$, $\ell, \ell' \in\{1, 2, 3, 4\}$, we deduce from Corollary  \ref{Cor-Compar-Gamma} that $\gamma_{\ell} \ge \gamma_{\ell'}$  whenever  $\varpi_{\ell} \ge \varpi_{\ell'}$.
%We see here that the investment risk still be of much borne by the  partners that contributed the more to the funding of the capital.  whom make the supra task of managing the project The more the investment is risky, the lower will be the supra profit ratio of partner $\ell$ with respect partner $\ell'$. This is because he exposes his labour or expertise contribution to the management to a loss when the investment is risky
\end{enumerate}
\item Suppose now that $(\kappa_1, \kappa_2,\kappa_3,\kappa_4) = (\frac{1}{8}, \frac{3}{8},\frac{1}{8},\frac{3}{8}) = (12.5\%, 37.5\%, 12.5\%,37.5\%)$. In this case, we have:
\begin{enumerate} [leftmargin=0.3cm,itemsep=0cm]
\item[$\diamond$]  If   $(c_1, c_2,c_3,c_4)=(1,1,1,1)$, then   $(\varpi_{1},$ $ \varpi_{2},$ $\varpi_{3},$ $\varpi_{4})=$ $(1/4, $ $1/4, $ $1/4, $ $1/4)$ and
\begin{eqnarray*}
&& (\gamma_1, \gamma_2, \gamma_3, \gamma_4)  \ = \  (23\%,\ 27 \%,\ 23\%, \ 27\%) \qquad \mbox{ if }\, \rho(R_T) =1/8 \\
&\mbox{ and }& (\gamma_1, \gamma_2, \gamma_3, \gamma_4)  \ =  \  (17\%,\ 33 \%,\ 17\%, \ 33 \%) \qquad \mbox{ if } \,\rho(R_T) =2/3. 
\end{eqnarray*}
Since  $ \varpi_{\ell} = \varpi_{\ell'}$, for all $\ell \not= \ell'$, $\ell, \ell' \in\{1, 2, 3, 4\}$, then  $\gamma_{\ell} \ge \gamma_{\ell'}$  whenever $\kappa_{\ell}  \ge  \kappa_{\ell'}$ (see Corollary  \ref{Cor-Compar-Gamma}).
\item[$\diamond$]  If $(c_1, c_2,c_3,c_4)=(1,2,1,4)$, then   $(\varpi_{1},$ $ \varpi_{2},$ $\varpi_{3},$ $\varpi_{4})=$ $\frac{1}{11} \, (4, $ $2, $ $4, $ $1)$ and 
\begin{eqnarray*}
&& (\gamma_1, \gamma_2, \gamma_3, \gamma_4)  \ = \  (33\%,\ 21 \%,\ 33\%, \ 13\%) \qquad \mbox{ if } \rho(R_T) =1/8 \\
&\mbox{ and }& (\gamma_1, \gamma_2, \gamma_3, \gamma_4)  \ \simeq \  (20\%,\ 32 \%,\ 20\%, \ 28 \%) \qquad \mbox{ if } \rho(R_T) =2/3. 
\end{eqnarray*}
We remark from the foregoing that   $\rho(R_T) =1/8$ and  $\gamma_1 \ge \gamma_4$ whereas $\kappa_1 \le \kappa_4$.  This is because  $\varpi_{1} - \varpi_{4}  > \kappa_4 - \kappa_1$ (see Corollary \ref{Cor-Compar-Gamma}).
\end{enumerate}
\item Now suppose that the company's manager (lthe general partner) is Partner 4  and that this one is bound to the entrepreneurs (shareholders) by a {\em PL-sharing mudarabah} contract. He does not contribute capital, but brings his expertise to help grow the business. Assume that $(\kappa_1, \kappa_2,\kappa_3) = (1/3, 1/3, 1/3)$ and $\rho(R_T,L) =1/2$.
\begin{enumerate} [leftmargin=0.3cm,itemsep=0cm]
\item[$\diamond$]  If  $(c_1, c_2,c_3,c_4)=(1,1,1,1)$, then  $(\varpi_{1},$ $ \varpi_{2},$ $\varpi_{3},$ $\varpi_{4})=$ $(1/4, $ $1/4, $ $1/4, $ $1/4)$ and  $(\gamma_1,$ $ \gamma_2,$ $ \gamma_3,$ $ \gamma_4)  $ $\ = \  $ $(29\%,$ $\ 29 \%,$ $\ 29\%, $ $\ 13\%)$. 
\item[$\diamond$] If $(c_1, c_2,c_3,c_4)=(2,2,2,3)$, then   $(\varpi_{1},$ $ \varpi_{2},$ $\varpi_{3},$ $\varpi_{4})=$ $ \frac{1}{9}(2, $ $2, $ $2, $ $3)$  and  $(\gamma_1,$  $ \gamma_2, $  $\gamma_3,$    $ \gamma_4)   =$  $   (28\%,$  $28 \%, $  $28\%, $   $ 16\%)$. 
\end{enumerate}
\end{enumerate}
}
\end{example}

 \begin{remark} (About a {\em diminishing musharakah} contract) Consider a viable $T$-term and $R$-asset-based {\em musharakah} contract. Equation \eqref{EqPart1-d-Prop2}  provides the profit-sharing ratios of the stockholders of the {\em musharakah} contract at time t = 0, so we develop a one-period model in this work. 

Consider a {\em diminishing musharakah} contract in which some stockholders wish to sell a portion of their stocks at time $t$, where $0 < t < T$, and suppose these stocks can be traded on a secondary stock market. In this situation, we need to  evaluate the profit-sharing ratios at time $t$, a problem which we propose to investigate separately from the present paper, both because of the need for more sophisticated tools from modern financial modelling, such as backward stochastic differential equations, and to avoid making the paper any longer.

%n fact,  consider a diminishing musharakah contract in which some stockholders wish to sell a portion of their stocks at time $t$, where $0 < t < T$, and suppose that these stocks can be traded on a secondary stock market. In this situation, we need to  evaluate the profit-sharing ratios at time $t$. We propose to investigate this problem separately to the present paper because both of the  need of more sophisticated tools coming from modern financial modelling like Backward Stochastic Differential Equations and to avoid making the paper any longer. 

%First, note that if we consider a viable $T$-term and $R$-asset-based {\em musharakah} contract,  Equation \eqref{EqPart1-d-Prop2} provides the profit-sharing ratios of the 'stockholders' of the {\em musharakah} contract at time $t = 0$. Now suppose that we have a secondary stock market in which stocks issued from a {\em musharakah} contract may be traded, or consider a {\em diminishing musharakah} contract in which some stockholders wish to sell a portion of their stocks at time $t$, where $0 < t < T$. In this situation, we need to price the stocks and evaluate the profit-sharing ratios at time $t$. This is a very challenging question for the islamic finance  industry. 
 \end{remark}

Next, we consider  a generalisation of  Proposition \ref{Prop-Musharakah-d-Partners2Parts} 2. to $d$ partners, where the $(d-1)$ partners of a {\em PL sharing - musharakah} contract  appoint a manager (the $d$'th partner) outside of their group  and enter into  a {\em wakalah} (an agency) contract with them.  

 \begin{proposition}  \label{Prop-Mush-Mudh-Wakalah} Consider  a  viable   $T$-term and $R$-asset based \,  {\em  PL - musharakah} contract involving  $d-1$ partners. Suppose the partners  appoint an external  manager  (partner $d$) and enter into  a {\em wakalah} contract with him,   which provides a fixed remuneration (a amount $p$ that is paid out  $k$ times up to the maturity $T$).  If the partners  agree  to rate their respective expected payoff by the coefficients $1 / c_{\ell}$, $\ell \in \{1, \ldots,  d\}$ then the $(c_1,\ldots,c_{d})$-fair profit sharing  ratios and periodic remuneration $p$ satisfy
 \begin{eqnarray}  
  \theta(r,T,k) \, p &=&    \varpi_{d}    \, \Delta(R_T,L)\\
 \gamma_{\ell}  &=& \Big( \frac{ \varpi_{d} }{d-1} +  \varpi_{\ell}\Big) \,  \big(1-\rho(R_T,L) \big)   + \kappa_{\ell} \, \rho(R_T,L), \ \ell = 1, \ldots, d-1. \quad   \label{EqPart1-Wakala-d-Prop1}  
\end{eqnarray}
Here,  $\theta(r,T,k) =  (1+r )^{ \mbox{\tiny $\frac{T}{k}$} } -1 \, / \, (1+r )^{ \mbox{\tiny $T$} } - 1$,  and  the weights  $(\varpi_{\ell})_{\{1 \le \ell \le d \}}$ is still defined  by $ \varpi_{\ell}  = c_{\ell} \, /  \, (  c_1 + \ldots + c_d )$, for every $\ell  \in \{1, \ldots, d\}$.

Furthermore, the expected payoff of each partner $\ell$ (even for the manager) is given by
\begin{equation}
\mbox{Pay}_{\ell} (\gamma_{\ell},\kappa_{\ell})  = \varpi_{\ell}\, \Delta(R_T,L),  \qquad \ell =1, \ldots,d.
\end{equation}
%
%\begin{eqnarray}
%  p  &=&  \,   \Delta(R_T,L), \\
%   \gamma_{\ell} & =& \varpi_{\ell}\, (1- \rho(R_T,L)) +  \kappa_{\ell}  \,  \rho(R_T,L)\\
%   &=&  \varpi_{\ell} \, \beta(R_T,L) +  \kappa_{\ell}  \,  \rho(R_T,L),
% \end{eqnarray}
%  where $\bar{\ell} = \{1,2\} \backslash \{ \ell \}$ and, for $\ell \in \{1, 2\}$, 
%  \[
%  c_{\ell,\bar {\ell}}  =   \frac{c_{ \bar{\ell} }}{c_1 c_2 + c_1 c_3 + c_2 c_3}  \Big( \frac{c_{\ell}}{2}  + c_3\Big).
%  \]

 \end{proposition}

Before presenting the proof, it is worth noting  that  \ $\gamma_1 + \ldots + \gamma_{d-1} =1$.  A typical example  where Proposition \ref{Prop-Mush-Mudh-Wakalah} applies is the situation where we have three partners: 
\begin{enumerate} 
\item[$\diamond$] an investor (Partner 1)  who provides all the capital  ($\kappa_1=1$),  
\item[$\diamond$] a bank (Partner 2)  acting  as an  agent  for the investor via a {\em wakalah} contract and arranges a fixed remuneration  $p$ that is paid out  $k$ times up to the maturity $T$; and
\item[$\diamond$] a company (Partner 3) in need of funding, bound to the bank   by a {\em PL-mudarabah} contract  ($\kappa_3=0$).
\end{enumerate}
%\textcolor{blue}{Proposition \ref{Prop-Mush-Mudh-Wakalah} also applies to an interbank system in which the central bank (Partner 2) acts as an agent for a commercial bank (Partner 1) and  invests the bank's  excess liquidity in a short-term investment offered by another bank (Partner 3) via a {\em PL-mudarabah} contract.}  

\begin{proof}  The $(c_1, \ldots, c_d)$-fair assumption implies for any $\ell=1, \ldots, d-1$,
 \begin{equation}  \label{Eq-Wakalah-gen-proof}
\mbox{Pay}_{\ell} (\gamma_{\ell},\kappa_{\ell},p,k)  \ / \ c_{\ell} =  \mbox{Pay}_3 (p,k) \ / \ c_d,
 \end{equation}  
where 
\begin{eqnarray}
\mbox{Pay}_{\ell} (\gamma_{\ell},\kappa_{\ell},p,k)   &=&   (1+r)^{\mbox{\tiny $-T$}}\, \big(\gamma_{\ell} \,  \mathbb E (R_T-L)^{+}  - \kappa_{\ell}\, \mathbb E (L-R_T)^{+} \big)  - \frac{\mbox{Pay}_3 (p,k)}{ d-1} \qquad   \label{Eq-Payoff-Wakalah-general1} \\
\mbox{ and } \quad \mbox{Pay}_3 (p,k) &=& p  \, \big(1+r \big)^{ \mbox{\tiny $-T$} }  \, \theta(r,T,k), \quad  \mbox{for  all } \, \ell=1, \ldots, d-1.  \nonumber 
\end{eqnarray}
It follows  from \eqref{Eq-Wakalah-gen-proof} that   for any $\ell=1, \ldots, d-1$,
\[
 \Big(\gamma_{\ell} \,  \mathbb E (R_T-L)^{+}  - \kappa_{\ell}\, \mathbb E (L-R_T)^{+}   -  \frac{p   \, \theta(r,T,k)}{d-1}  \Big) \ / \  c_{\ell}  = p   \, \theta(r,T,k)  \ / \ c_d.
\]
Set,  for all  $\ell, d \in \{1, \ldots, d\}$,  $\mathcal{ P}_d = \{1, \ldots, d\}$  and  $\bar {\mathcal{P}}_{d}(\ell) = \mathcal{P}_d  \backslash \{\ell\}$.  %multipliant les deux membres de l'\'egalit\'e pr\'ec\'edente par $ \prod_{i \in \bar\mathcal{P}_{d-1}(\ell)} (1/ c_i) $, on obtient:
Multiplying both sides of the previous equation by $ \prod_{i \in \bar{\mathcal{P}}_{d-1}(\ell)} (1/ c_i) $  gives
\[
 \Big(\gamma_{\ell} \,  \mathbb E (R_T-L)^{+}  - \kappa_{\ell}\, \mathbb E (L-R_T)^{+}   -  \frac{p   \, \theta(r,T,k)}{d-1}  \Big)  \prod_{i=1}^{d-1}  c_i^{-1}   =   p   \, \theta(r,T,k) \prod_{i \in \bar{\mathcal{P}}_{d}(\ell)} c_i^{-1}.
\]
Since $\gamma_1+ \ldots +\gamma_{d-1}=1$ and $\kappa_1 + \ldots + \kappa_{d-1}=1$, taking the sum over the $\ell$'s (from $1$ to $d-1$) yields:
\[
\Big(\Delta(R_T,L) - p   \, \theta(r,T,k)  \Big)  \prod_{i \in \mathcal{P}_{d-1}} c_i ^{-1}  = p   \, \theta(r,T,k)  \sum_{\ell \in \mathcal{P}_{d-1}} \prod_{i \in \bar{\mathcal{P}}_{d}(\ell)} c_i^{-1}.
\]
We deduce that 
\begin{eqnarray*}
\Delta(R_T,L)   \prod_{i \in \mathcal{P}_{d-1}} c_i^{-1}  &=& p   \, \theta(r,T,k)  \Big(  \prod_{i \in \mathcal{P}_{d-1}} c_i ^{-1} +  \sum_{\ell \in \mathcal{P}_{d-1}} \prod_{i \in \bar{\mathcal{P}}_{d}(\ell)} c_i^{-1} \Big) \\
&=& p   \, \theta(r,T,k)  \Big(  \prod_{i \in \bar{\mathcal{P}}_{d} (d)} c_i^{-1}  +  \sum_{\ell \in \mathcal{P}_{d-1}} \prod_{i \in \bar{\mathcal{P}}_{d}(\ell)} c_i^{-1} \Big) \\
&=&  p   \, \theta(r,T,k)  \Big(  \sum_{\ell \in \mathcal{P}_{d}} \prod_{i \in \bar{\mathcal{P}}_{d}(\ell)} c_i^{-1} \Big),
\end{eqnarray*}
and, finally, that,
\[
\theta(r,T,k) \, p =    \varpi_d   \, \Delta(R_T,L).  
\]
Now, replacing $ \theta(r,T,k) \, p $ by its value in Equation \eqref{Eq-Wakalah-gen-proof} gives 
\[
c_{\ell}\, \Big(\gamma_{\ell} \,  \mathbb E (R_T-L)^{+}  - \kappa_{\ell}\, \mathbb E (L-R_T)^{+}   -  \frac{\varpi_{d} \, \Delta(R_T,L)}{d-1}  \Big)  = c_d \,p   \, \theta(r,T,k).\]
This implies that
\begin{eqnarray*}
\gamma_{\ell} & = & \Big( \frac{ \varpi_{d}  }{d-1} + \frac{c_{\ell}}{c_d} \varpi_{d} \Big) \,  \big(1-\rho(R_T,L) \big)   + \kappa_{\ell} \, \rho(R_T,L) \\
&=&\Big( \frac{\varpi_{d} }{d-1} + \varpi_{\ell} \Big) \,  \big(1-\rho(R_T,L) \big)   + \kappa_{\ell} \, \rho(R_T,L).
\end{eqnarray*}
Now, the expected payoff of  each partner $\ell$ is obtained by replacing $ \theta(r,T,k) \, p$ and $\gamma_{\ell}$ by their values in Equation \eqref{Eq-Wakalah-gen-proof}
\end{proof}
\begin{example}  \label{Exam-mudha-wakalah}
Consider the following situation involving three partners. The first partner is an investor who provides all the capital ($\kappa_1=1$). The second partner is a bank acting as an agent for the investor via a {\em wakalah} contract, arranging fixed remuneration $p$ to be paid $k$ times up to maturity ($T$). The third partner is a company requiring funding, bound to the bank by a {\em PL-mudarabah} contract ($\kappa_3=0$).

%Consider the following situation involving  three partners. The first partner is an investor who provides all the capital (so that $\kappa_1=1$). The second partner is a bank that acts as an agent for the investor via a {\em wakalah} contract, arranging a fixed remuneration $p$ that is paid $k$ times up to maturity ($T$). The third partner is a company that requires funding and is bound to the bank by a {\em PL-mudarabah} contract ($\kappa_3=0$).

Suppose  they invest an amount $L$ in company asset $R$ for a maturity period $T$.  Given  the vector of  weighting coefficients  $c = (c_1, c_2, c_3)$,  the respective shares of the bank,  the investor and the company are given by:
  \begin{eqnarray*}  
  \theta(r,T,k) \, p &=&    \varpi_{3}     \, \Delta(R_T,L),\\
 \gamma_{1}  &=& \Big( \frac{\varpi_{3}}{2} + \varpi_{1}\Big) \,  \big(1-\rho(R_T,L) \big)   +  \rho(R_T,L), \\  
 \gamma_{2}  &=& \Big( \frac{\varpi _{3}}{2} + \varpi_{2} \Big). 
\end{eqnarray*}
For example, if $c = (1,1,1)$ and $\rho (R_T, L) = 1/2$, then
\[
 \theta(r,T,k)  p = \frac{1}{3}  \Delta(R_T,L), \qquad \gamma_1 = \frac{3}{4}, \qquad \gamma_{2} = \frac{1}{4}.
\]
This  leads to the expected investment profit  being sharing equally between the three partners, i.e. $\frac{1}{3}  \Delta(R_T,L)$ each.

\end{example}

\subsection{How it applies to central banking?}

Consider an interbank investment market in which banks can create and sell securities or investment notes backed by the financing of major government or corporate projects. These instruments can be structured under various types of contract, including  {\em PL-sharing-style joint venture agreements}. If a bank is unable to raise the necessary funds to finance a project, it can   turn to the central bank. The central bank can then use deposits from other banks - which it is already mandated to invest in such projects (then, the central bank acts as an agent)  -  to invest under a  {\em PL-sharing-style joint venture} agreement. The central bank may also invest under a {\em PL-sharing-style limited partnership}  agreement when necessary.

We know that controlling the money supply is one of the central bank's missions. The question arises as to how this can be achieved in an interbank system based on an interbank investment market. This paper provides an answer regarding the money supply in relation to investment. Indeed, the central bank can adjust its monetary policy by setting a minimum profit-sharing rate ({\small MPS} rate), possibly sector-specific depending on its monetary policy guidelines, on {\em PL-sharing-style joint venture} contracts. This allows it to implement (we denote by the {\small FPS} rate the $c$-fair profit-sharing rate on {PL-sharing-style joint venture} contracts, including  {\em PL-sharing-style limited partnership}   contracts):

\begin{enumerate}
\item Either {\em a macroeconomic stimulus policy through a reduction in the {\small MPS} rate}: when economic activity slows down, the central bank can set the {\small MPS} rate below the {\small FPS} rate to try to revitalize it by encouraging innovation and investment. Thus, commercial banks will reduce their margin on investments to the benefit of businesses, which can pass this reduction on to prices. This will increase household purchasing power.
\item {\em A deterrent macroeconomic policy through an increase in the {\small MPS} rate}: when investment is too high, risking deflation, the central bank can set the {\small MP} rate $>$ {\small FPS} rate to discourage investment and encourage consumption by economic agents, in order to rebalance supply and demand levels.
\end{enumerate}

%Consider an interbank investment market in which banks can create securities or investment notes backed by the financing of major government or corporate projects, to sell them on this market. These instruments can be structured under various types of contracts, including \textcolor{blue}{{\em PL-sharing-style joint venture}} agreements. When a bank is unable to raise the necessary funds to finance a project, it can turn to the central bank. The central bank can then use deposits from other banks - which it is already mandated to invest in such projects - to invest under a \textcolor{blue}{{\em PL-sharing-style joint venture}}  agreement, for example. The central bank may also, when necessary, invest under a \textcolor{blue}{{\em PL-sharing-style limited partnership}}   agreement.

\section{Modelisation perspectives}

 First, note that following our approach, the computation of the profit-sharing ratios and  all the also  quantities of interest depend on both  $\rho(R_T,L)$ and $\Delta(R_T,L)$.  In practice, to compute these quantities, we need to specify an adequate model (dependent on the observations) for the dynamics of the investment asset $R$.  Among the possible models, we may enumerate the usual econometrics models, such as autoregressive (AR) models, moving average  (MA) models, autoregressive integrated moving average  (ARIMA) models,  autoregressive conditional heteroskedasticity (ARCH) models, as well as  their generalizations, such as  GARCH models.  We may also use diffusion model governed   by stochastic differential equations, as used  in finance.  Of these finance-based models, we  will consider    the famous Black-Scholes model,  for which all the quantities of interest can be computed explicitly.

\subsection{The Black-Scholes model}

\subsubsection{The model and the main result} In the Black-Scholes model, we  suppose that the investment asset  $(R_t)_{t \in [0,T]}$ evolves   following the SDE:     
\begin{equation} \label{Eq-Stochastic} 
\mbox{d}R_{t}  = \mu\, R_t \,\mbox{d}t  + \sigma   R_t \,\mbox{d}W_t, \qquad  R_0 = x_0, \quad t \in [0,T],
\end{equation}
where  $\mu$ is the instantaneous return, $\sigma$ is the instantaneous volatility, and,  the process  $(W_t)_{t \in [0,T]}$ is a standard brownian motion defined on a probability space $(\Omega, \mathcal {F}, \mathbb P)$. 

We have the following easy result.

 \begin{proposition}  If the dynamics of $R$ is that of Black-Scholes, then, an investment on the asset $R$ is viable if and only if  $\mu>0$. Furthermore, we have  
\begin{eqnarray} 
&&  \rho(R_T,L)  = -  \frac{\Phi(- \vartheta(\mu, \sigma)) - e^{\mu T} \Phi(- \vartheta(\mu, \sigma) - \sigma \sqrt{T})}{\Phi(\vartheta(\mu, \sigma)) - e^{\mu T} \Phi(\vartheta(\mu, \sigma) + \sigma \sqrt{T})}  \\
 &\mbox{ and }& \Delta (R_T,L)  = L (e^{\mu T} -1),    \label{Eq-Delta-BS}
 \end{eqnarray}
where   $ \vartheta(\mu, \sigma)  = (2\mu -\sigma^2)T\, / \, (2 \sigma \sqrt{T})$  and  $\Phi(\cdot)$  is the cdf of the standard gaussian distribution.
 \end{proposition}

 \begin{proof} The viability result follows from \eqref{Eq-Delta-BS}.  In fact
  \[
  \Delta(R_T,L)>0  \quad \Longleftrightarrow \quad \mu>0.
  \]
  For the rest, we just need to show that
  $ \mathbb E ((L-R_T)^{+})    $ $= L \big(\Phi(- \vartheta(\mu, \sigma)) - e^{\mu T} \Phi(- \vartheta(\mu, \sigma) - \sigma \sqrt{T})\big)
  $ and $  \mathbb E ((R_T - L)^{+})   =  L \big( \Phi(\vartheta(\mu, \sigma)) - e^{\mu T} \Phi(\vartheta(\mu, \sigma) + \sigma \sqrt{T}) \big)$,  and use the relation $ \Delta(R_T,L) =    \mathbb E ((R_T - L)^{+}) -   \mathbb E ((L-R_T)^{+}).$
  \end{proof}

\subsubsection{The investment risk  as  functions of the volatility and  the return}  
\begin{figure}[htpb]
 \begin{center}
 % \!\includegraphics{Risque-Fonct-Sig.pdf}	
   \!\includegraphics[width=11.0cm,height=7cm]{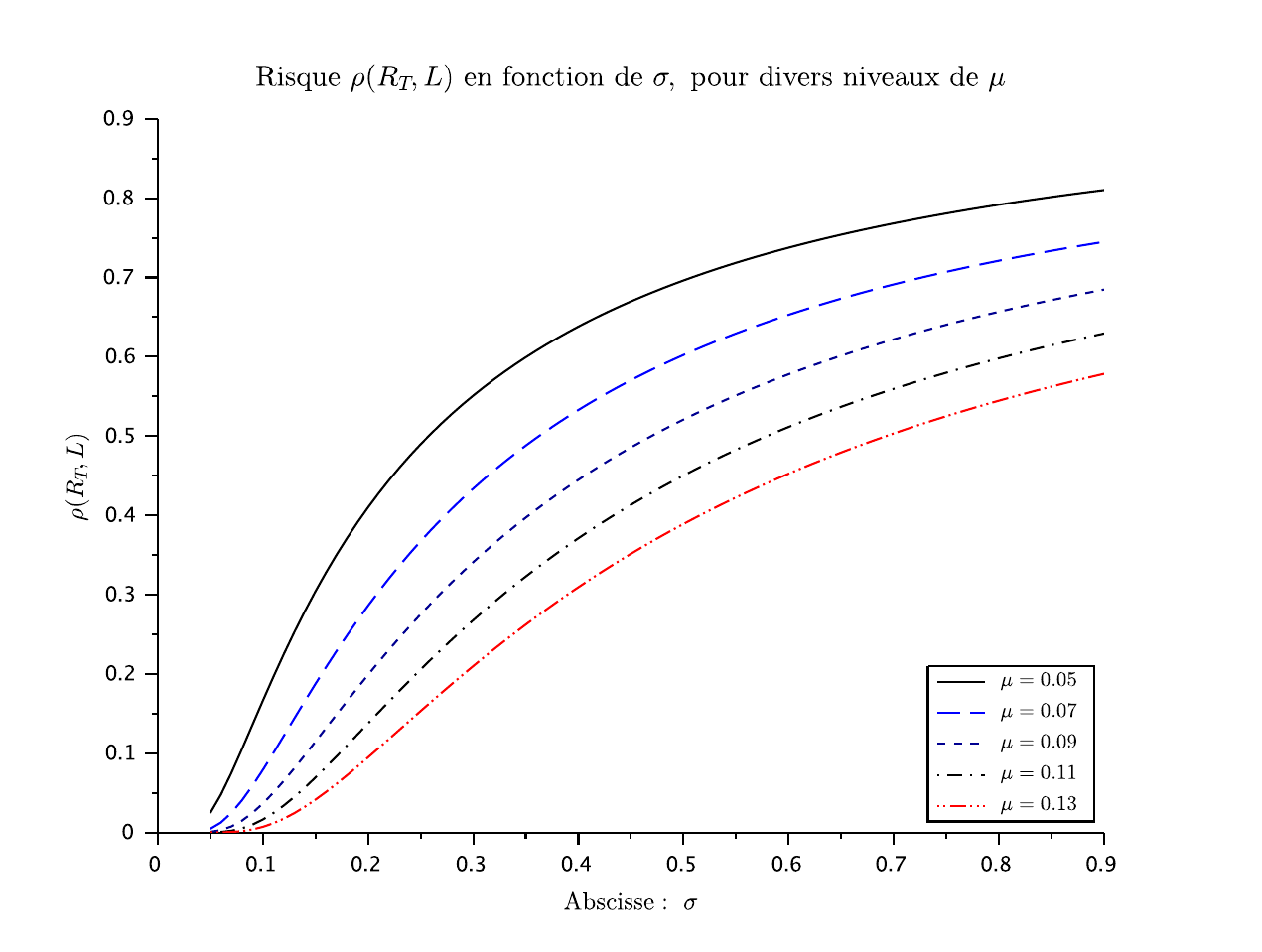}	
%\hfill   \!\includegraphics[width=7.0cm,height=5.0cm]{BS-Mult-Path-sig005.pdf}
  \caption{\footnotesize  {\it Black-Scholes model}: $\rho(R_T,L)$  as a function of the volatility $\sigma$, for different values of  $\mu$.}
  \label{fig-Rho-fct-Sig}
  \end{center}
\end{figure}

To understand how the investment risk $\rho(R_T,L)$ behaves with respect to volatility and return parameters ($\sigma$ and  $\mu$), Figure \ref{fig-Rho-fct-Sig} shows how  $\rho(R_T,L)$  varies with  $\sigma$ and  $\mu$:  with $\sigma$ ranging from $5\%$ to $90\%$ in steps of $1\%$, and $\mu \ \in \{0.05, 0.07, 0.09, 0.11, 0.13 \}$.

As expected,  we see that 
\begin{enumerate}
\item[$\diamond$]  for a fixed level of return,  the investment risk is a nondecreasing function of the volatility: the greater  the volatility of the investment asset, the high is the investment risk;
\item[$\diamond$]  for a fixed level of volatility, the investment risk is a non-increasing function of the return: the greater the yield, the smaller it is.
\end{enumerate}

\subsubsection{Practical use of the previous results}
Consider a bank that aims to invest in a specific sector of the economy (fintechs, for example) through a {\em mudarabah} contract with a maturity period of $T$. Suppose the bank wants to capture two-fifths of the sector's development potential. The following are some steps the bank can take to achieve this:

\begin{enumerate}
\item Assume that we  have access to a dataset for the investment sector and that we can model the dynamics of the investment asset $R$ up to maturity $T$.
\item  Ensure that  the investment is  viable by checking  (from the model)  that  $\Delta(R_T,L)>0$ and  compute (or approximate) $\rho(R_T,L)$.
\item Since  we want to capture   two-fifths  of the  $\Delta(R_T,L)$, we may invest the amount  $L$ in a large number of independent projects (say, more than $30$, for example,  bearing  in mind the law of large number). Now, if for example,
\begin{eqnarray*}
& &   \rho(R_T,L)=1/4,  \mbox{ we negotiate a } \gamma_1 =55\%  \mbox{ share of the profits from each contract}. \\
&&  \rho(R_T,L)=1/2,  \mbox{ we negotiate a } \gamma_1 =70\%  \mbox{ share of the profits from each contract}. 
%
%Si $\rho(R_T,L)=1/2$, on n\'egocie une part $\gamma_1 =70\%$ des profits  avec chaque contrat. Ceci  nous assure d'avoir les $2/5$ de $\Delta(R_T,L)$.
\end{eqnarray*}
%Both previous strategies guarantee to hold $2/5$ of   $\Delta(R_T,L)$.
\end{enumerate}

\section{Conclusion} In this work, we introduce a new concept of a $c$-fair profit-sharing ratio, which is based on Islamic investment contracts such as  {\em musharakah} or {\em mudarabah}, and can be  combined with a {\em wakalah} contract.  The $c$-fair concept allocates profits based on each partner's contribution to the overall success of the undertaking.    

We demonstrate that the profit-sharing ratio is proportional to investment risk and opportunity, weighted by capital and labour contributions, respectively.

To calculate the quantities of interest, we must model the dynamics of the investment asset.  The feasibility of this step depends on the nature of the available dataset and must be thoroughly investigated for any given dataset. Of the possible models used in finance and econometrics, we considered the well-known  Black-Scholes model, for which all the quantities of interest can be computed explicitly. This model also enables us  to observe how investment risk is expected to behave: it  increases with volatility and decreases with return.

Finally, this work sheds new light on these Islamic contracts and on the engineering and industry of Islamic finance. It may also help Islamic finance institutions  value Islamic investment contracts and related financial instruments such as Sukuk instruments based on these  contracts, as well as manage  {\em Takaful} (Islamic insurance) via a {\em wakalah} or  {\em wakalah}-{\em moudharabah} models.

\bigskip
\section*{Acknowledgement} The author would like to thank  Stephen Koffler for his relevant  and useful  review of the manuscript.

% \newpage
%  \end{comment}
 % \cite{Lessy, Askari-Iqbal-Mirakhor, Ayub}
\bibliographystyle{plain}
%\bibliography{Bibliographie}

\end{document}